\documentclass{article}
\pdfoutput=1
\usepackage{bbding}
\usepackage{graphicx}
\usepackage{afterpage}
\usepackage[caption=false]{subfig}
\usepackage{multirow}
\usepackage{multicol}
\usepackage{array}
\usepackage{xcolor}
\usepackage{float}
\usepackage[normalem]{ulem}
\usepackage{listings}
\usepackage{makecell}
\usepackage[hyphens]{url}
\usepackage{hyperref}
\usepackage{ulem}
\usepackage{tablefootnote}
\usepackage{flushend}
\usepackage{authblk}

\title{A Machine Learning Approach to Online Fault Classification in HPC Systems}

\date{}
\author[1,2]{Alessio Netti}
\author[2]{Zeynep Kiziltan}
\author[2]{Ozalp Babaoglu}
\author[4]{Alina S\^irbu}
\author[3]{Andrea Bartolini}
\author[3]{Andrea Borghesi}
\affil[1]{Leibniz Supercomputing Centre, Garching bei M\"unchen, Germany}
\affil[2]{Department of Computer Science and Engineering, University of Bologna,
Italy}
\affil[3]{Department of Electrical, Electronic and Information Engineering,
University of Bologna, Italy}
\affil[4]{Department of Computer Science, University of Pisa, Italy}

\affil[ ]{\textit {alessio.netti@lrz.de,alina.sirbu@unipi.it}}
\affil[ ]{\textit {\{zeynep.kiziltan,ozalp.babaoglu,a.bartolini,andrea.borghesi3\}@unibo.it}}

\begin{document}

\maketitle

\begin{abstract}
As High-Performance Computing (HPC) systems strive towards the \emph{exascale} goal, failure rates both at the hardware and software levels will increase significantly. Thus, detecting and classifying faults in HPC systems as they occur and initiating corrective actions before they can transform into failures becomes essential for continued operation. Central to this objective is fault injection, which is the deliberate triggering of faults in a system so as to observe their behavior in a controlled environment. In this paper, we propose a fault classification method for HPC systems based on machine learning. The novelty of our approach rests with the fact that it can be operated on streamed data in an online manner, thus opening the possibility to devise and enact control actions on the target system in real-time. We introduce a high-level, easy-to-use fault injection tool called FINJ, with a focus on the management of complex experiments. In order to train and evaluate our machine learning classifiers, we inject faults to an in-house experimental HPC system using FINJ, and generate a fault dataset which we describe extensively. Both FINJ and the dataset are publicly available to facilitate resiliency research in the HPC systems field. Experimental results demonstrate that our approach allows almost perfect classification accuracy to be reached for different fault types with low computational overhead and minimal delay.
\end{abstract}

\section{Introduction}
\label{section:introduction}

\paragraph{Motivation}
Modern scientific discovery is increasingly being driven by computation~\cite{villa2014scaling}. In a growing number of areas where experimentation is either impossible, dangerous or costly, computing is often the only viable alternative towards confirming existing theories or devising new ones. As such, High-Performance Computing (HPC) systems have become fundamental ``instruments'' for driving scientific discovery and industrial competitiveness. Exascale ($10^{18}$ operations per second) is the moonshot for HPC systems. Reaching this goal is bound to produce significant advances in science and technology through higher-fidelity simulations, better predictive models and analysis of greater quantities of data, leading to vastly-improved manufacturing processes and breakthroughs in fundamental sciences ranging from particle physics to cosmology. Future HPC systems will achieve exascale performance through a combination of faster processors and massive parallelism~\cite{ashby2010opportunities}. 

With Moore's Law having reached its limit, the only viable path towards higher performance has to consider switching from increased transistor density towards increased core count, thus increased sockets count. This, however, presents major obstacles. With everything else being equal, the fault rate of a system is directly proportional to the number of sockets used in its construction~\cite{cappello2014toward}. But everything else is not equal: exascale HPC systems will also use advanced low-voltage technologies that are much more prone to aging effects~\cite{bergman2008exascale} together with system-level performance and power modulation techniques, such as dynamic voltage frequency scaling, all of which tend to increase fault rates~\cite{engelmann2017resilience}. Economic forces that push for building HPC systems out of commodity components aimed at mass markets only add to the likelihood of more frequent unmasked hardware faults. Finally, complex system software, often built using open-source components, to deal with more complex and heterogeneous hardware, fault masking and energy management, coupled with legacy applications will significantly increase the potential for faults~\cite{jones2012application}. It is estimated that large parallel jobs will encounter a wide range of failures as frequently as once every 30 minutes on exascale platforms~\cite{snir2014addressing}. At these rates, failures will prevent applications from making progress. Consequently, exascale performance, even if achieved nominally, cannot be sustained for the duration of most applications that often run for long periods.

To be usable in production environments with acceptable \emph{quality of service} levels, exascale systems need to improve their resiliency by several orders of magnitude. Therefore, future exascale HPC systems must include automated mechanisms for masking faults, or recovering from them, so that computations can continue with minimal disruptions. In our terminology, a \emph{fault} is defined as an anomalous behavior at the software or hardware level that can lead to illegal system states (\emph{errors}) and, in the worst case, to service interruptions (\emph{failures})~\cite{gainaru2015errors}. In this paper, we limit our attention to improving the resiliency of HPC systems through the use of mechanisms for detecting and classifying faults as soon as possible since they are the root causes of errors and failures. An important technique for this objective is \emph{fault injection}: the deliberate triggering of faults in a system so as to observe their behavior in a controlled environment, enable development of new prediction and response techniques as well as testing of existing ones~\cite{hsueh1997fault}. For fault injection to be effective, dedicated tools are necessary, allowing users to trigger complex and realistic fault scenarios in a reproducible manner. 

\paragraph{Contributions} 
The contributions of our work are several fold. 
First, we propose and evaluate a fault classification method based on supervised Machine Learning (ML) suitable for online deployment in HPC systems as part of an infrastructure for building mechanisms to increase their resiliency. Our approach relies on a collection of performance metrics that are readily available in most HPC systems. A novel aspect of our proposal is its ability to work online with live streamed data as opposed to traditional offline techniques that work with archived data. Our experimental results show that the method we propose can classify almost perfectly several types of faults, ranging from hardware malfunctions to software issues and bugs. In our method, classification can be achieved with little computational overhead and with minimal delay, making it suitable for online use. We characterize the performance of our proposed solution in a realistic online use scenario where live streamed data is fed to fault classifiers both for training and for detection, dealing with issues such as class imbalance and ambiguous system states. Most existing studies, on the contrary, consider offline scenarios and rely on extensive manipulation of archived data, which is not feasible in online scenarios. Furthermore, switching from an offline to an online approach based on streamed data opens up the possibility for devising and enacting control actions in real-time.

Second, we introduce an easy-to-use open-source Python fault injection tool called FINJ. A relevant feature of FINJ is the possibility of seamless integration with other injection tools targeted at specific fault types, thus enabling users to coordinate faults from different sources and different system levels. By using FINJ's \emph{workload} feature, users can also specify lists of applications to be executed and faults to be triggered on multiple nodes at specific times, with specific durations. FINJ thus represents a high-level, flexible tool, enabling users to perform complex and reproducible experiments, aimed at revealing the complex relations that may exist between faults, application behavior and the system itself. FINJ is also extremely easy to use: it can be set up and executed in a matter of minutes, and does not require the writing of additional code in most of its usage scenarios. To the best of our knowledge, FINJ is the first portable, open-source tool that allows users to perform and control complex injection experiments, that can be integrated with heterogeneous fault types and that includes workload support, while retaining ease of use and a quick setup time.

As a third and final contribution, our evaluation is based on a dataset consisting of monitoring data that we acquired from an experimental HPC system (called Antarex) by injecting faults using FINJ. We make the Antarex dataset publicly available and describe it extensively for use in the community. This is an important contribution to the HPC field, since commercial operators are very reluctant to share trace data containing information about faults in their HPC systems~\cite{kondo2010failure}.

\paragraph{Organization}
The rest of the paper is organized as follows. In the next section we put our work in context with respect to related work. In Section~\ref{section:architecture}, we introduce the FINJ tool and present a simple use case in Section~\ref{section:casestudy} to show how it can be deployed. In Section~\ref{section:workload}, we describe the Antarex dataset that will be used for evaluating our supervised ML classification models. In Section~\ref{section:features}, we discuss the features extracted from the Antarex dataset to train the classifiers and in Section~\ref{section:experimentalresults}, we present our experimental results. We conclude in Section~\ref{section:conclusions}.

\section{Related Work}
\label{section:relatedwork}

Fault injection for prediction and detection purposes is a recent topic of intense activity, and several studies have proposed tools with varying levels of abstraction. Calhoun et al.~\cite{calhoun2014flipit} devised a compiler-level fault injection tool focused on memory bit-flip errors, targeting HPC applications. De Bardeleben et al.~\cite{debardeleben2011experimental} proposed a logic error-oriented fault injection tool. This tool is designed to inject faults in virtual machines, by exploiting emulated machine instructions through the open-source virtual machine and processor emulator (QEMU). Both works focus on low-level fault-specific tools and do not provide functionality for the injection of complex workloads, and for the collection of produced data, if any.

Stott et al.~\cite{stott2000nftape} proposed NFTAPE, a high-level and generic tool for fault injection. This tool is designed to be integrated with other fault injection tools and triggers at various levels, allowing for the automation of long and complex experiments. The tool, however, has aged considerably, and is not publicly available. A similar fault injection tool was proposed by Naughton et al.~\cite{naughton2009fault}, which has never evolved beyond the prototype stage and is also not publicly available, to the best of our knowledge. Moreover, both tools require users to write a fair amount of wrapper and configuration code, resulting in a complex setup process. The Gremlins Python package\footnote{\url{https://github.com/toddlipcon/gremlins}} also supplies a high-level fault injector. However, it does not support workload or data collection functionalities, and experiments on multiple nodes cannot be performed.

Joshi et al.~\cite{joshi2011prefail} introduced the PREFAIL tool, which allows for the injection of failures at any code entry point in the underlying operating system. This tool, like NFTAPE, employs a coordinator process for the execution of complex experiments. It is targeted at a specific fault type (code-level errors) and does not permit performing experiments focused on performance degradation and interference. Similarly, the tool proposed by Gunawi et al.~\cite{gunawi2011fate}, named FATE, allows the execution of long experiments; furthermore, it is focused on reproducing specific fault sequences, simulating real scenarios. Like PREFAIL, it is limited to a specific fault type, namely I/O errors, thus greatly limiting its scope.

Automated fault detection and characterization through system performance metrics is a common application for fault injection, and for most of the tools discussed above. However, machine learning-based methods using fine-grained monitored data (i.e., sampling once per second) are more recent. Tuncer et al.~\cite{tuncer2018online} proposed a framework for the diagnosis of performance anomalies in HPC systems. Since they deal only with performance anomalies that result in longer runtimes for applications, they do not consider faults that lead to errors and failures, which cause a disruption in the computation. Moreover, the data used to build the test dataset was not acquired continuously, but rather in small chunks related to single application runs. Thus, it is not possible to determine the feasibility of this method when dealing with streamed, continuous data from an online HPC system. A similar work was proposed by Baseman et al.~\cite{baseman2016interpretable}, which focuses on identifying faults in HPC systems through temperature sensors. Ferreira et al.~\cite{ferreira2008characterizing} analyzed the impact of CPU interference on HPC applications by using a kernel-level noise injection framework. Both approaches deal with specific fault types, and are therefore limited in scope.

Other researchers have focused on using coarser-grained data (i.e., sampling once per minute) or on reducing the dimension of collected data, while retaining good detection accuracy. Bodik et al.~\cite{bodik2010fingerprinting} aggregated monitored data by using fingerprints, which are built from quantiles corresponding to different time epochs. Lan et al.~\cite{lan2010toward} discussed an outlier detection framework based on principal component analysis. Guan et al.~\cite{guan2012cda,guan2013adaptive} proposed works focused on finding the correlations between performance metrics and fault types through a most relevant principal components method. Wang et al.~\cite{wang2010online} proposed a similar entropy-based outlier detection framework suitable for use in online systems. These frameworks, which are very similar to threshold-based methods, are not suitable for detecting the complex relationships that may exist between different performance metrics under certain faults. One notable work in threshold-based fault detection is the one proposed by Cohen et al.~\cite{cohen2004correlating}, in which probabilistic models are used to estimate threshold values for performance metrics and detect outliers. This approach requires constant human intervention to tune thresholds, and lacks flexibility.

\section{The FINJ Tool}
\label{section:architecture}
In this section, we first discuss how fault injection is achieved in FINJ. We then present the architecture of FINJ and discuss its implementation. Customizing FINJ for different purposes is easy, thanks to its portable and modular nature. 

\subsection{Task-based Fault Injection}
\label{section:working}
Fault injection is achieved through \emph{tasks} that are executed on target nodes. Each task corresponds to either an HPC application or a fault-triggering program, and has a specification for its execution. As demonstrated by Stott et al.~\cite{stott2000nftape}, this approach allows for the integration in FINJ of any third-party fault injection framework that can be triggered externally. In any case, many fault-triggering programs are supplied with FINJ (see Section \ref{subsection:features}), allowing users to experiment with anomalies out-of-the-box.

A sequence of tasks defines a \emph{workload}, which is a succession of scheduled application and fault executions at specific times, reproducing a realistic working environment for the fault injection process. A task is specified by the following attributes:

\begin{itemize}
\item \emph{args}: the full shell command required to run the selected task. The command must refer to an executable file that can be accessed from the target hosts;
\item \emph{timestamp}: the time in seconds at which the task must be started, expressed as a relative offset;
\item \emph{duration}: the task's \emph{maximum allowed} duration, expressed in seconds, after which it will be abruptly terminated. This duration can serve as an \emph{exact} duration as well, with FINJ restarting the task if it finishes earlier, and terminating it if it lasts more. This behavior depends on the FINJ configuration (see Section~\ref{section:component});
\item \emph{isFault}: defines whether the task corresponds to a fault-triggering program, or to an application;
\item \emph{seqNum}: a sequence number used to uniquely identify the task inside a workload;
\item \emph{cores}: an optional attribute which is the list of CPU cores that the task is allowed to use on target nodes, enforced through a \emph{NUMA Control} policy \cite{lameter2013numa}.
\end{itemize}

A workload is stored in a \emph{workload file}, which contains the specifications of all the tasks of a workload in CSV format. The starting time of each task is expressed as a relative offset, in seconds, with respect to the first task in the workload. A particular execution of a given workload then constitutes an \emph{injection session}.

\begin{figure}[t]
  \centering
	\includegraphics[width=0.4\textwidth]{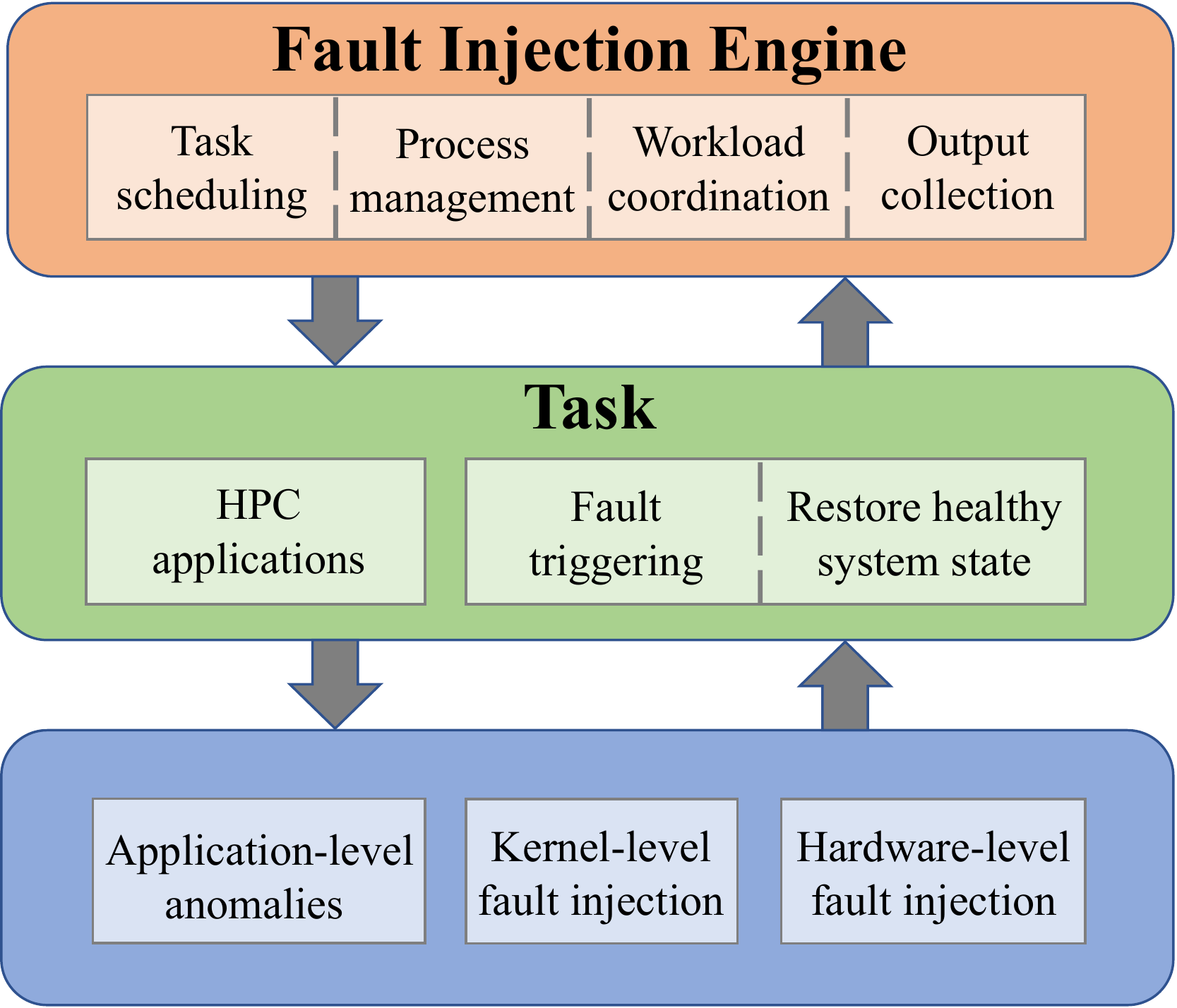}
  \caption{A diagram representing the responsibility of each entity involved in the fault injection process.}
  \label{fig:taskdiagram}
\end{figure} 

In our approach, the responsibility of each entity involved in the fault injection process is isolated, as depicted in Figure~\ref{fig:taskdiagram}. The fault injection engine of FINJ manages the execution of tasks on target nodes and the collection of their output. Tasks contain all the necessary logic to run applications or to trigger any other low-level fault injection framework (e.g., by means of a writable file or system call). At the last level, the activated low-level fault injector handles the actual triggering of faults, which can be at the software, kernel, or even hardware level. 

\subsection{FINJ Architecture}
\label{section:archoverview}

\begin{figure}[t]
  \centering
	\includegraphics[width=0.45\textwidth]{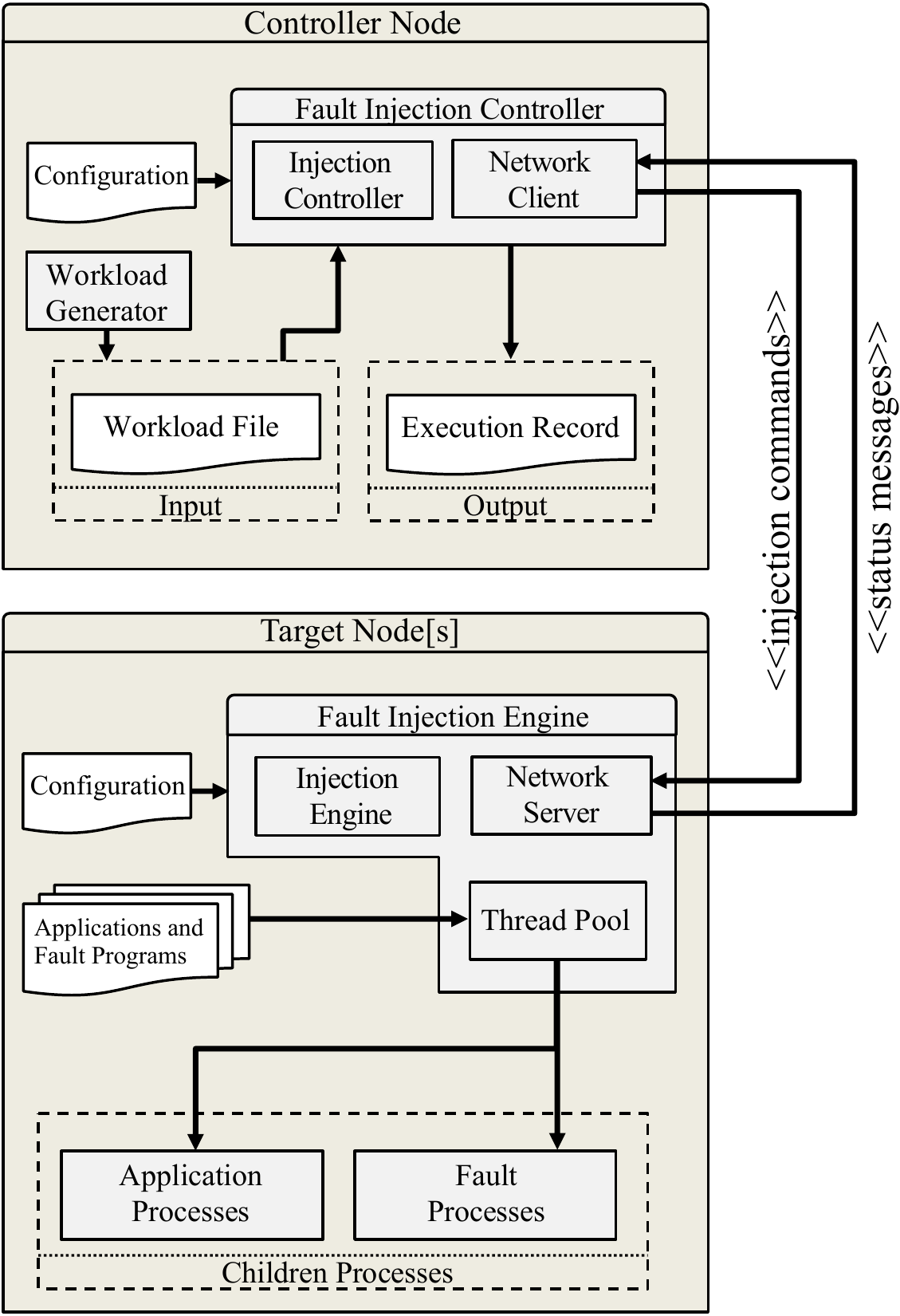}
  \caption{Architecture of the FINJ tool showing the division between a controller node (top) and a target node (bottom).}
  \label{fig:architecture}
\end{figure} 

FINJ consists of a \emph{fault injection controller} and a \emph{fault injection engine}, which are designed to run on separate machines. The high-level structure of the FINJ architecture is illustrated in Figure~\ref{fig:architecture}.

The \emph{controller} orchestrates the injection process, and should be run on an external node that is not affected by the faults. The controller maintains connections to all nodes involved in the injection session, which run fault injection engine instances and whose addresses are specified by users. Therefore, injection sessions can be performed on multiple nodes at the same time. The controller reads the task entries from the selected workload file. For each task the controller sends a command to all target hosts, instructing them to start the new task at the specified time. Finally, the controller collects and stores all status messages (e.g., task start and termination, status changes) produced by the target hosts. 

The \emph{engine} is a daemon running on nodes where faults are to be injected. The engine waits for task commands to be received from remote controller instances. Engines can be connected to multiple controllers at the same time, however task commands are accepted from one controller at a time, the \emph{master} of the injection session. The engine manages received task commands by assigning them to a dedicated thread from a \emph{pool}. The thread manages all aspects related to the execution of the task, such as spawning the necessary subprocesses and sending status messages to controllers. Whenever a fault causes a target node to crash and reboot, controllers are able to re-establish and recover the previous injection session, given that the engine is set up to be executed at boot time on the target node.

\subsection{Architecture Components}
\label{section:component}
FINJ is based on a highly modular architecture, and therefore it is very easy to customize its single components in order to add or tune features.

\paragraph{Network} Engine and controller instances communicate through a network layer, and communication is achieved through a simple message-based protocol. Specifically, we implemented \emph{client} and \emph{server} software components for the exchange of messages. Fault injection controllers use a client instance in order to connect to fault injection engines, which in turn run a server instance which listens for incoming connections. A message can be either a \emph{command} sent by a controller, related to a single task, or a \emph{status} message, sent by an engine, related to status changes in its machine. All messages are in the form of \emph{dictionaries}. This component also handles resiliency features such as automatic re-connection from clients to servers, since temporary connection losses are to be expected in a fault injection context.

\begin{figure}[t]
  \centering
	\includegraphics[width=0.45\textwidth]{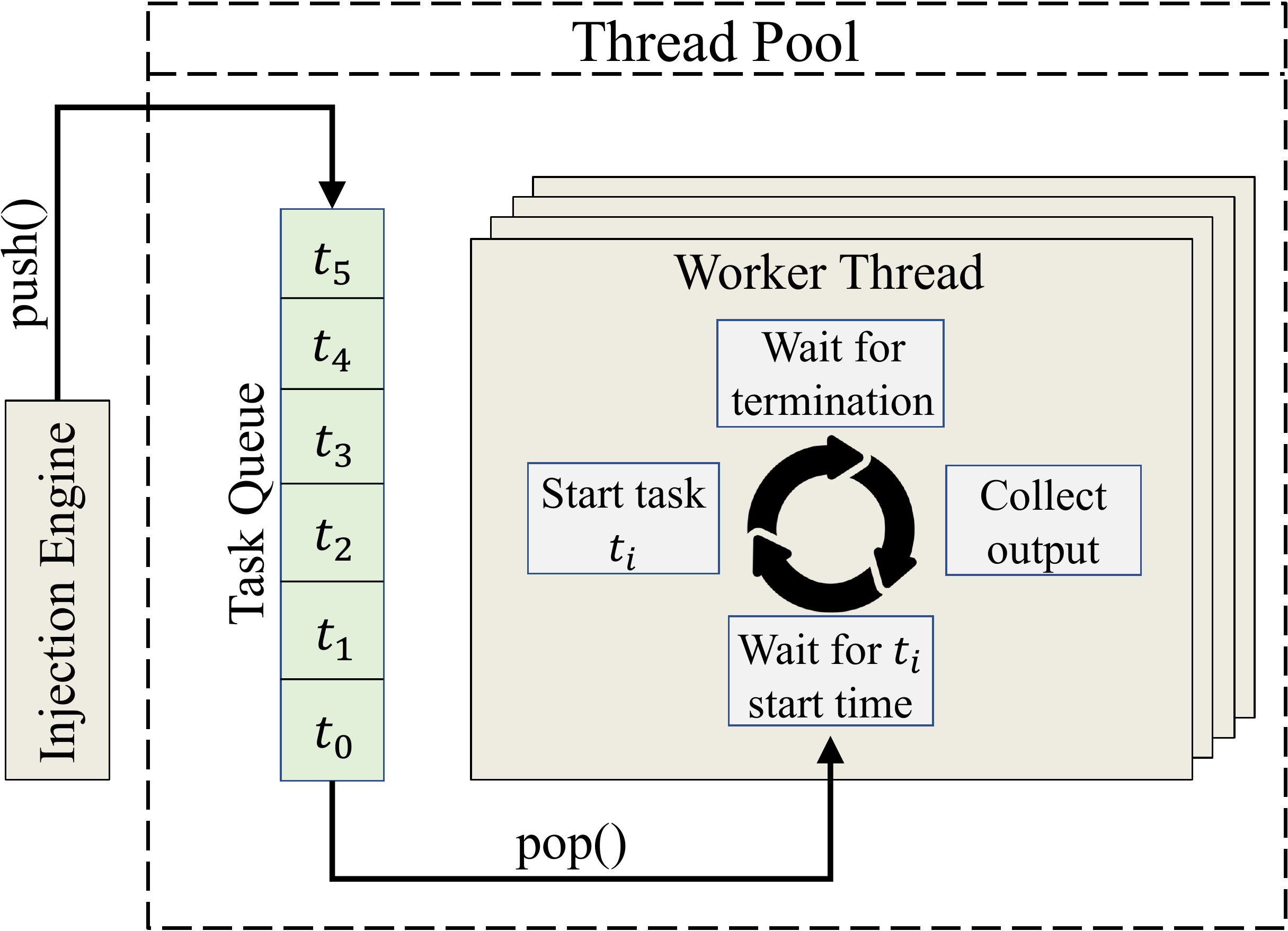}
  \caption{A representation of the life cycle of a task, as managed by the thread pool of the fault injection engine.}
  \label{fig:pooldiagram}
\end{figure}

\paragraph{Thread Pool} Task commands in the engines are assigned to a thread in a pool as they are received. Each thread manages all aspects of a task assigned to it. Specifically, the thread sleeps until the scheduled starting time of the task (according to its time-stamp); then, it spawns a subprocess running the specified task, and sends a message to all connected controllers to inform them of the event. At this point, the thread waits for the task's termination, depending on its duration and on the current configuration. Finally, the thread sends a new status message to all connected hosts informing them of the task's termination, and returns to sleep. The amount of threads in the pool, which is a configurable parameter, determines the maximum number of tasks that can be executed concurrently. Since threads in the pool are started only once during the engine's initialization, and wake up for minimal amounts of time when a task needs to be started or terminated, we expect their impact on the performance to be negligible. The life cycle of a task, as managed by a worker thread, is represented in Figure~\ref{fig:pooldiagram}.

\paragraph{Input and Output} The input and output of all data related to injection sessions are performed by controller instances, and are handled by \emph{reader} and \emph{writer} entities. By default, the input/output format is CSV, which was chosen due to its extreme simplicity and generality. \emph{Input} is constituted by \emph{workload} files, that include one entry for each injection task, as described in Section~\ref{section:working}. \emph{Output}, instead, is made up of two parts: I) the \emph{execution log}, which contains entries corresponding to status changes in the target nodes (e.g., start and termination, encountered errors and connection loss or recovery events); II) output produced by single tasks.

\paragraph{Configuration} FINJ's runtime behavior can be customized by means of a configuration file. This file includes several options that alter the behavior of either the controller or engine instances. Among the basic options, it is possible to specify the listening TCP port for engine instances, and the list of addresses of target hosts, to which controller instances should connect at launch time. The latter is useful when injection sessions must be performed on large sets of nodes, whose addresses can be conveniently stored in a file. More complex options are also available. For instance, it is possible to define a series of commands corresponding to tasks that must be launched together with FINJ, and must be terminated with it. This option proves especially useful when users wish to set up monitoring frameworks, such as the \emph{Lightweight Distributed Metric Service} (LDMS)~\cite{agelastos2014lightweight}, to be launched together with FINJ in order to collect system performance metrics during injection sessions.

\paragraph{Workload Generation} While writing workload files manually is possible, this is time-consuming and not desirable for long injection sessions. Therefore, we implemented in FINJ a \emph{workload generation} tool, which can be used to automatically generate workload files with certain statistical features, while trying to combine flexibility and ease of use. The workload generation process is controlled by three parameters: a maximum \emph{time span} for the total duration of the workload expressed in seconds, a statistical distribution for the \emph{duration} of tasks, and another one for their \emph{inter-arrival} times. We define the inter-arrival time as the interval between the start of two consecutive tasks. These distributions are separated in two sets, for fault and application tasks, thus amounting to a total of four. They can be either specified analytically by the user or can be fitted from real data, thus reproducing realistic behavior.

A workload is composed as a series of fault and application tasks that are selected from a list of possible shell commands. To control the composition of workloads, users can optionally associate to each command a probability for its selection during the generation process, and a list of CPU cores for its execution, as explained in Section~\ref{section:working}. By default, commands are picked uniformly. Having defined its parameters, the workload generation process is then fairly simple. Tasks are randomly generated in order to achieve statistical features close to those specified as input, and are written to an output CSV file, until the maximum imposed time span is reached. Alongside the full workload, a \emph{probe} file is also produced, which contains one entry for each task type, all with a short fixed duration, representing a lightweight workload version. This file can be used during the setup phase to test the correct configuration of the system, making sure that all tasks are correctly found and executed on the target hosts, without having to run the entire heavy workload. 

\subsection{Implementation}
\label{section:implementation}

FINJ is implemented in Python, an object-oriented, high-level interpreted programming language\footnote{\url{https://www.python.org/events/python-events/}}, and can be used on all major operating systems. All dependencies are included in the Python distribution, and the only optional external dependency is the \emph{scipy} package, which is needed for the workload generation functionality. The source code is publicly available on GitHub\footnote{\url{https://github.com/AlessioNetti/fault\_injector}} under the MIT license, together with its documentation, usage examples and several fault-triggering programs. FINJ works on Python versions 3.4 and above.

\begin{figure}[t]
  \centering
	\includegraphics[width=0.48\textwidth]{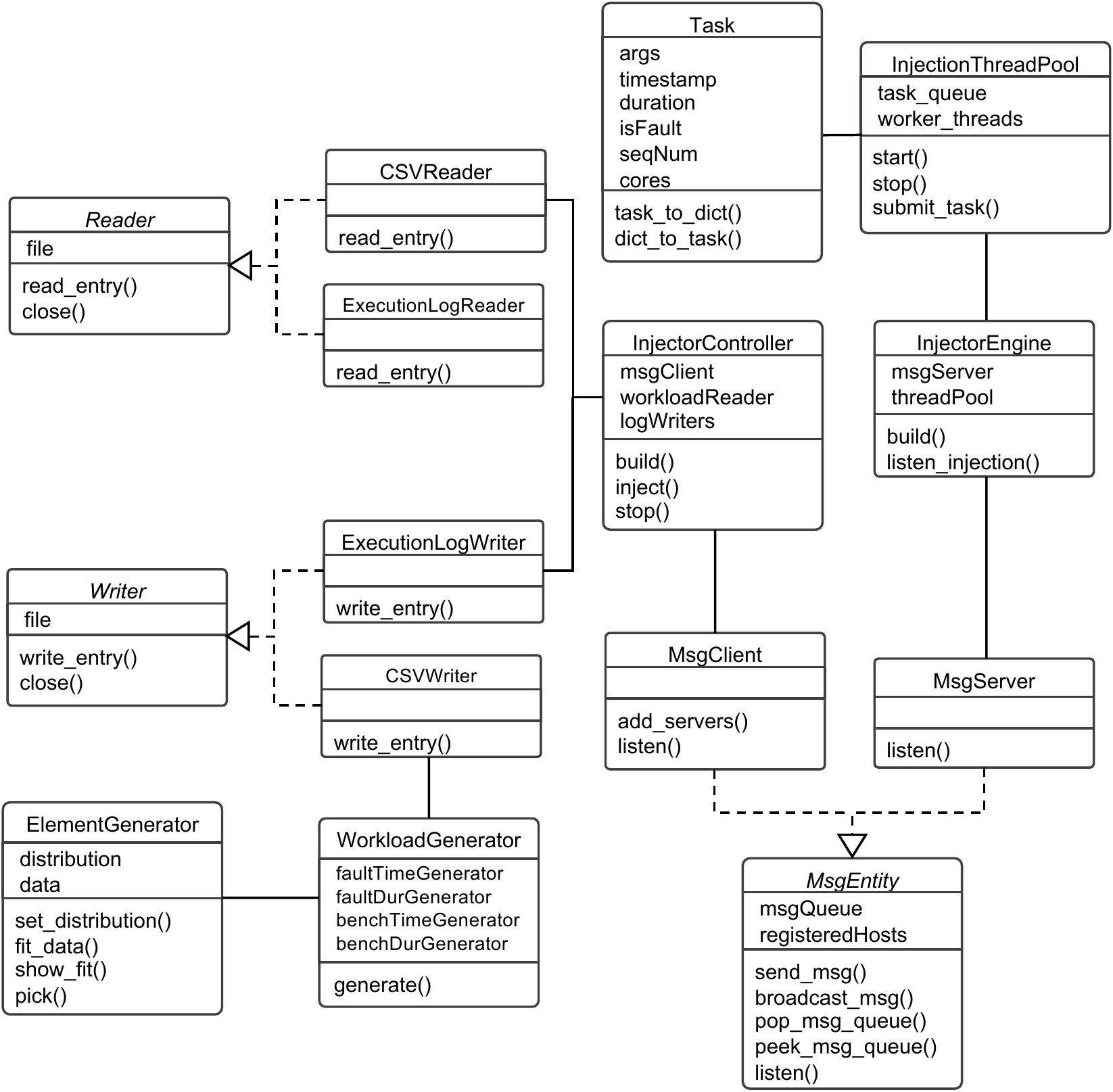}
  \caption{The class diagram of FINJ.}
  \label{fig:classdiagram}
\end{figure} 

In Figure \ref{fig:classdiagram}, we illustrate the class diagram of FINJ. The fault injection engine and the fault injection controller are represented by the \emph{InjectorEngine} and \emph{InjectorController} classes. Users can instantiate these classes and start injection sessions directly, by using the \emph{listen\_injection} method to put the engine in listening mode, and the \emph{inject} method of the controller, which allows to start the injection session itself. Scripts are supplied with FINJ to create controller and engine instances from a command-line interface, simplifying the process. The \emph{InjectionThreadPool} class, instead, supplies the thread pool implementation used to execute and manage the tasks.

The network layer of the tool is represented by the \emph{MsgClient} and \emph{MsgServer} classes, which implement the message and queue-based client and server used for communication. Both classes are implementations of the \emph{MsgEntity} abstract class, which provides the interface for sending and receiving messages, and implements the basic mechanisms that regulate the access to the underlying queue.

Input and output are instead handled by the \emph{Reader} and \emph{Writer} abstract classes and their implementations: \emph{CSVReader} and \emph{CSVWriter} handle the reading and writing of workload files, while \emph{ExecutionLogReader} and \emph{ExecutionLogWriter} handle the execution logs generated by injection sessions. Since these classes are all implementations of abstract interfaces, it is easy for users to customize them for different formats. Tasks are modeled by the \emph{Task} class that contains all attributes specified in Section~\ref{section:working}.

Lastly, access to the workload generator is provided through the \emph{WorkloadGenerator} class, which is the interface used to set up and start the generation process. This class is backed by the \emph{ElementGenerator} class, which offers basic functionality for fitting data and generating random values. This class acts as a wrapper on scipy's \emph{rv\_continuous} class, which generates random variables.

\section{Using FINJ}
\label{section:casestudy}

In this section we demonstrate the flow of execution of FINJ through a concrete example carried out on a real HPC node, and provide insight on its overhead.

\subsection{Sample Execution}
\label{section:execution}

Here we consider a sample fault injection session. The employed CSV workload file is illustrated in Figure~\ref{fig:inputsample}. The experimental setup for this test, both for fault injection and monitoring, is the same as that for the acquisition of the Antarex dataset, which we present in Section~\ref{section:workload}. Python scripts are supplied with FINJ to start and configure \emph{engine} and \emph{controller} instances: their usage is explained on the GitHub repository of the tool, together with all configuration options.

In this workload, the first task corresponds to an HPC application and is the Intel Distribution for the well-known \emph{High-Performance Linpack} (HPL) benchmark, optimized for Intel Xeon CPUs. This task starts at time 0 in the workload, and has a maximum allowed duration of roughly 30 minutes. The following two tasks are fault-triggering programs: \emph{cpufreq} which dynamically reduces the maximum allowed CPU frequency, emulating performance degradation, and \emph{leak} which creates a memory leak in the system, eventually using all available RAM. The cpufreq program requires appropriate permissions, so that users can access the files controlling Linux CPU governors. Both fault programs are discussed in detail in Section~\ref{section:workload}. The HPL benchmark is run with 8 threads, pinned on the first 8 cores of the machine, while the cpufreq and leak tasks are forced to run on cores 6 and 4, respectively. Note also that the tasks must be available at the specified path on the systems running the engine, which in this case is relative to the location of the FINJ launching script. 

Having defined the workload, the injection engine and controller must be started. For this experiment, we run both on the same machine. The controller instance will then connect to the engine and start executing the workload, storing all output in a CSV file which is unique for each target host. Each entry in this file represents a status change event, which in this case is the start or termination of tasks, and is flagged with its absolute time-stamp on the target host. In addition, any errors that were encountered are also reported. When the workload is finished, the controller terminates. It can be clearly seen from this example how easily a FINJ experiment can be configured and run on multiple CPU cores.

\begin{figure}[t]
\begin{lstlisting}[frame=tb]
timestamp;duration;seqNum;isFault;cores;args 
0;1723;1;False;0-7;./hpl lininput
355;244;2;True;6;sudo ./cpufreq 258
914;291;3;True;4;./leak 316
\end{lstlisting}
\caption{A sample FINJ workload file.}
\label{fig:inputsample}
\end{figure}

\begin{figure}[t]
  \centering
	\includegraphics[width=0.48\textwidth]{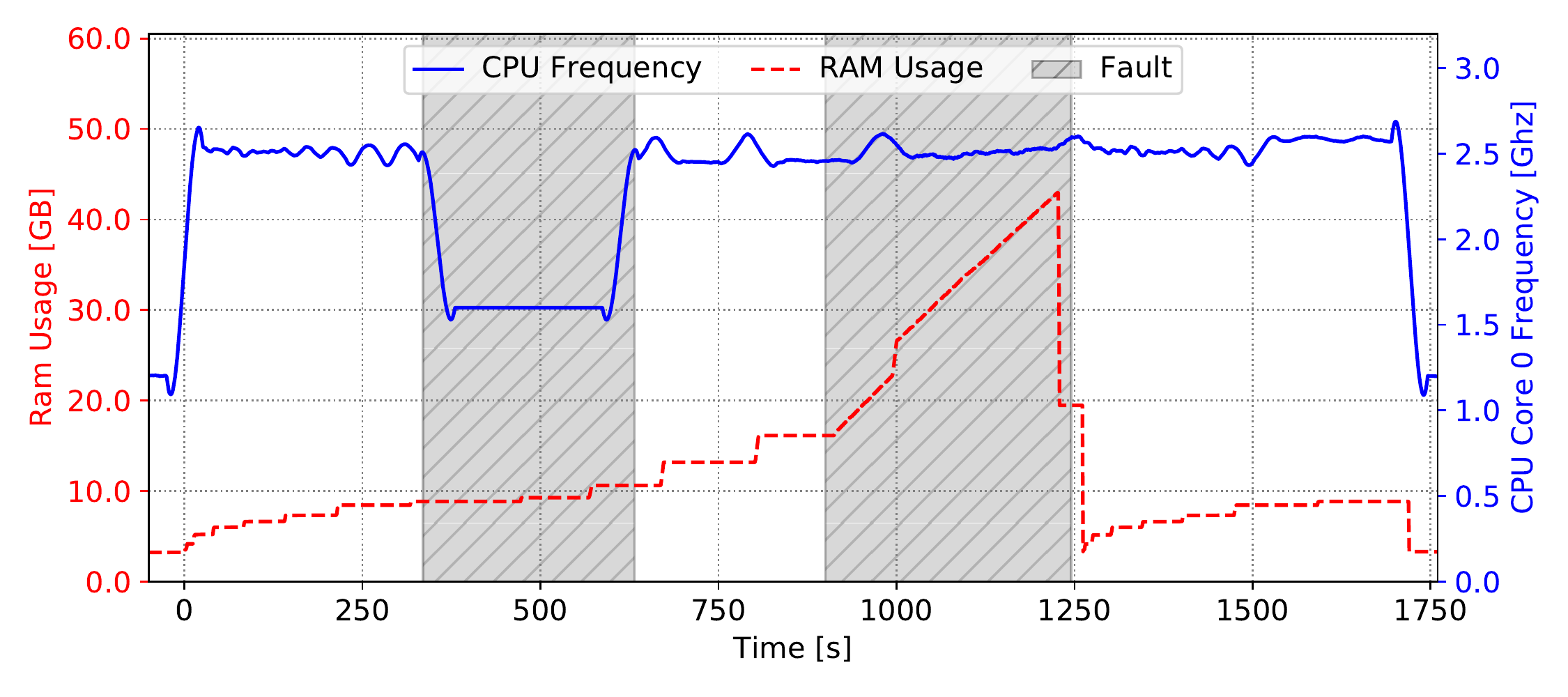}
  \caption{CPU Frequency and RAM Usage, as monitored on the target system during a sample FINJ injection session.}
  \label{fig:execution}
\end{figure}

At this point, the data generated by FINJ can be easily compared with other data, for example performance metrics collected through a monitoring framework, in order to better understand the system's behavior under faults. In Figure \ref{fig:execution} we show the total RAM usage and the CPU frequency of core 0, as monitored by the LDMS framework. The benchmark's profile is simple, showing a constant CPU frequency while RAM usage slowly increases as the application performs tests on increasing matrix sizes. The effect of our fault programs, marked in gray, can be clearly observed in the system. The \emph{cpufreq} fault causes a sudden drop in CPU frequency, resulting in reduced performance and longer computation times, while the \emph{leak} fault causes a steady, linear increase in RAM usage. Even though saturation of the available RAM is not reached, this peculiar behavior can be used for prediction purposes.

\subsection{Overhead of FINJ}
\label{section:overhead}
We performed tests in order to evaluate the overhead that FINJ may introduce. To do so, we employed the same machine used in Section~\ref{section:workload} together with the HPL benchmark, this time configured to use all 16 cores of the machine. We executed the HPL benchmark 20 times directly on the machine by using a shell script, and then repeated the same process by embedding the benchmark in 20 tasks of a FINJ workload file. FINJ was once again instantiated locally. In both conditions the HPL benchmark scored an average running time of roughly 320 seconds, therefore leading us to conclude that the impact of FINJ on running applications is negligible. This was expected, since FINJ is designed to perform only the bare minimum amount of operations in order to start and manage tasks, without affecting their execution.

\section{The Antarex Dataset}
\label{section:workload}
The dataset contains trace data that we collected from an HPC system (called Antarex) located at ETH Zurich while we injected faults using FINJ. The dataset is publicly available for use by the community and all the details regarding the test environment, as well as the employed applications and faults are extensively documented.\footnote{\url{https://zenodo.org/record/2553224}} In this section, we give a comprehensive overview of the dataset.

\subsection{Dataset Acquisition and Experimental Setup}
To acquire data, we executed some HPC applications and at the same time injected faults using FINJ in a single compute node of Antarex. We acquired the data in four steps by using four different FINJ workload files. Each data acquisition step consists of application and fault program runs related to either the CPU and memory components, or the hard drive, either in single-core or multi-core mode. This resulted in 4 blocks of nearly 20GB and 32 days of data in total, whose structure is summarized in Table~\ref{table:datasetstructure}. We acquired the data by continuous streaming, thus any study based on it will easily be reproducible on a real HPC system, in an online way.

We acquired data from a single HPC compute node for several reasons. First, we focus on fault detection models that operate on a per-node basis and that do not require knowledge about the behavior of other nodes. Second, we assume that the behavior of different compute nodes under the same faults will be comparable, thus a model generated for one node will be usable for other nodes as well. Third, our fault injection experiments required alterations to the Linux kernel of the compute node or special permissions. These are possible in a test environment, but not in a production one, rendering the acquisition of a large-scale dataset on multiple nodes impractical.

\begin{table}[t!]
\centering
\renewcommand{\arraystretch}{1.5} 
{
\caption{A summary of the structure for the Antarex dataset.}
\label{table:datasetstructure}
\fontsize{8.5}{8.5}\selectfont
%\begin{tabular}{c{1.5cm}|c{1.5cm}|c{1cm}|c{1cm}|c{1cm}}
\begin{tabular}{c|c|c|c|c}
 & \makecell{\textbf{Block} \\ \textbf{I}} & \makecell{\textbf{Block} \\ \textbf{III}} & \makecell{\textbf{Block} \\ \textbf{II}} &  \makecell{\textbf{Block} \\ \textbf{IV}} \\ \hline
 \textbf{Type} & \multicolumn{2}{c|}{CPU-Mem} & \multicolumn{2}{c}{HDD} \\ \hline
 \textbf{Parallel} & No & Yes & No & Yes \\ \hline
 \textbf{Duration} & \multicolumn{2}{c|}{12 days} & \multicolumn{2}{c}{4 days} \\ \hline
 \textbf{Applications} & \multicolumn{2}{c|}{\makecell{\quad \\ DGEMM~\cite{dongarra1990algorithm}, HPCC~\cite{luszczek2006hpc}, \\ STREAM~\cite{stream}, HPL~\cite{dongarra2003linpack} \\ \quad}} & \multicolumn{2}{c}{\makecell{IOZone~\cite{iozone}, \\ Bonnie++~\cite{bonnie}}} \\ \hline
 \textbf{Faults} & \multicolumn{2}{c|}{\makecell{\quad \\ leak, memeater, ddot, \\ dial, cpufreq, pagefail \\ \quad}} & \multicolumn{2}{c}{\makecell{ioerr,  copy}} \\
\end{tabular}
}
\end{table}

The Antarex compute node is equipped with two Intel Xeon E5-2630 v3 CPUs, 128GB of RAM, a Seagate ST1000NM0055-1V4 1TB hard drive and runs the CentOS 7.3 operating system. The node has a default Tier-1 computing system configuration. We used FINJ in a Python 3.4 environment. We also used the LDMS monitoring framework to collect performance metrics, so as to create features for fault classification purposes, as described in Section~\ref{section:features}. We configured LDMS to sample several metrics at each second, which come from the following seven different plug-ins:

\begin{enumerate}
\item \emph{meminfo} collects general information on RAM usage;
\item \emph{perfevent} collects CPU performance counters;
\item \emph{procinterrupts} collects information on hardware and software interrupts;
\item \emph{procdiskstats} collects statistics on hard drive usage;
\item \emph{procsensors} collects sensor data related to CPU temperature and frequency;
\item \emph{procstat} collects general metrics about CPU usage;
\item \emph{vmstat} collects information on virtual memory usage.
\end{enumerate}

This configuration resulted in a total of 2094 metrics to be collected each second. Some of the metrics are node-level, and describe the status of the compute node as a whole, others instead are core-specific and describe the status of one of the 16 available CPU cores. In order to minimize noise and bias in the sampled data, we chose to analyze, execute applications and inject faults into only 8 of the 16 cores available in the system, and therefore used only one CPU. On the other CPU of the system, instead, we executed the FINJ and LDMS tools, which rendered their CPU overhead negligible.

\subsection{FINJ Workload}
\label{subsection:features}

\begin{figure}[t!]
 \centering
 \captionsetup[subfigure]{}
  \subfloat[Histogram of fault durations.]{
    \includegraphics[width=0.42\textwidth,trim={0 5 10 5}, clip=true]{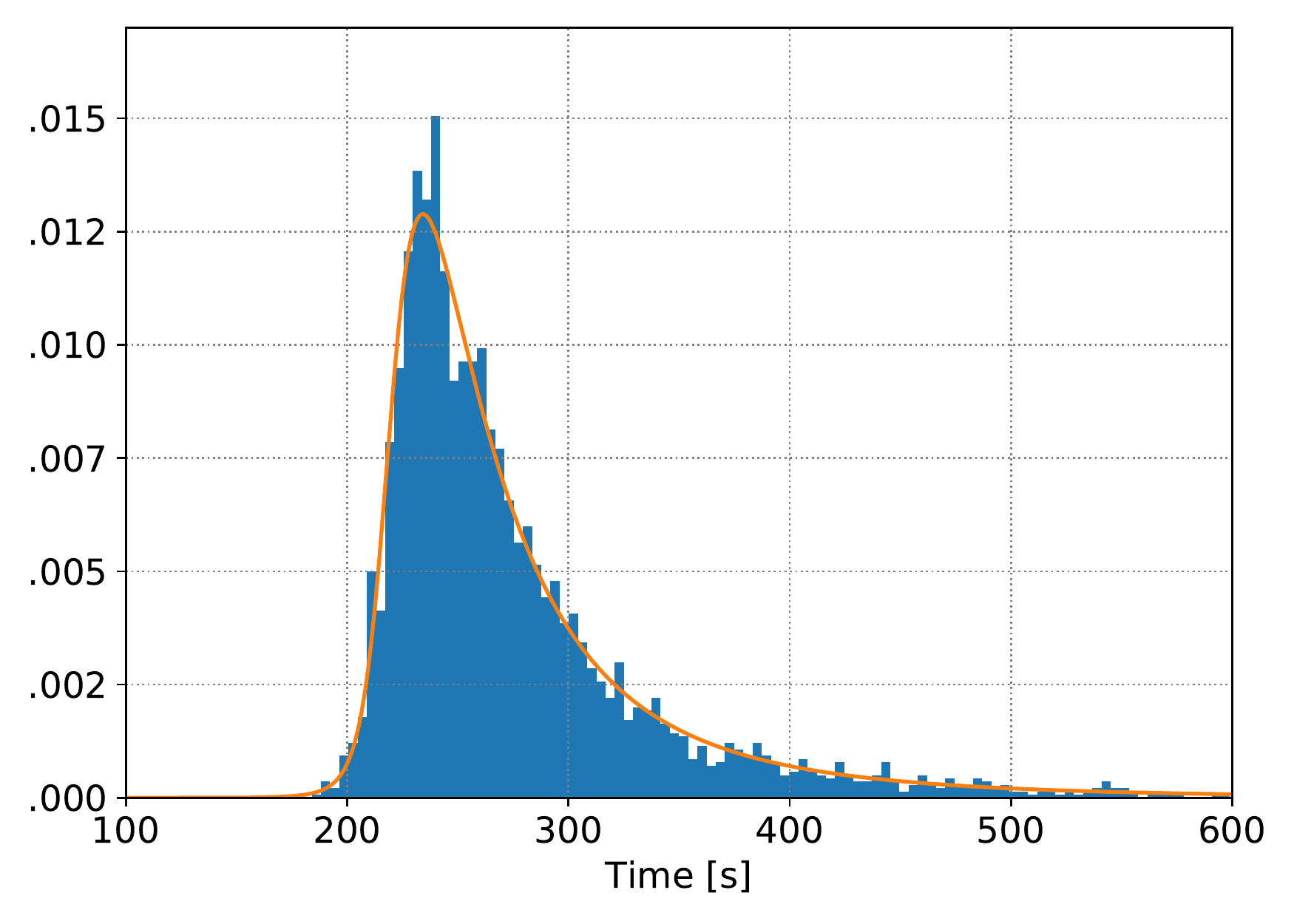}
  	} \\
  \subfloat[Histogram of fault inter-arrival times.]{
    \includegraphics[width=0.42\textwidth,trim={0 5 10 5}, clip=true]{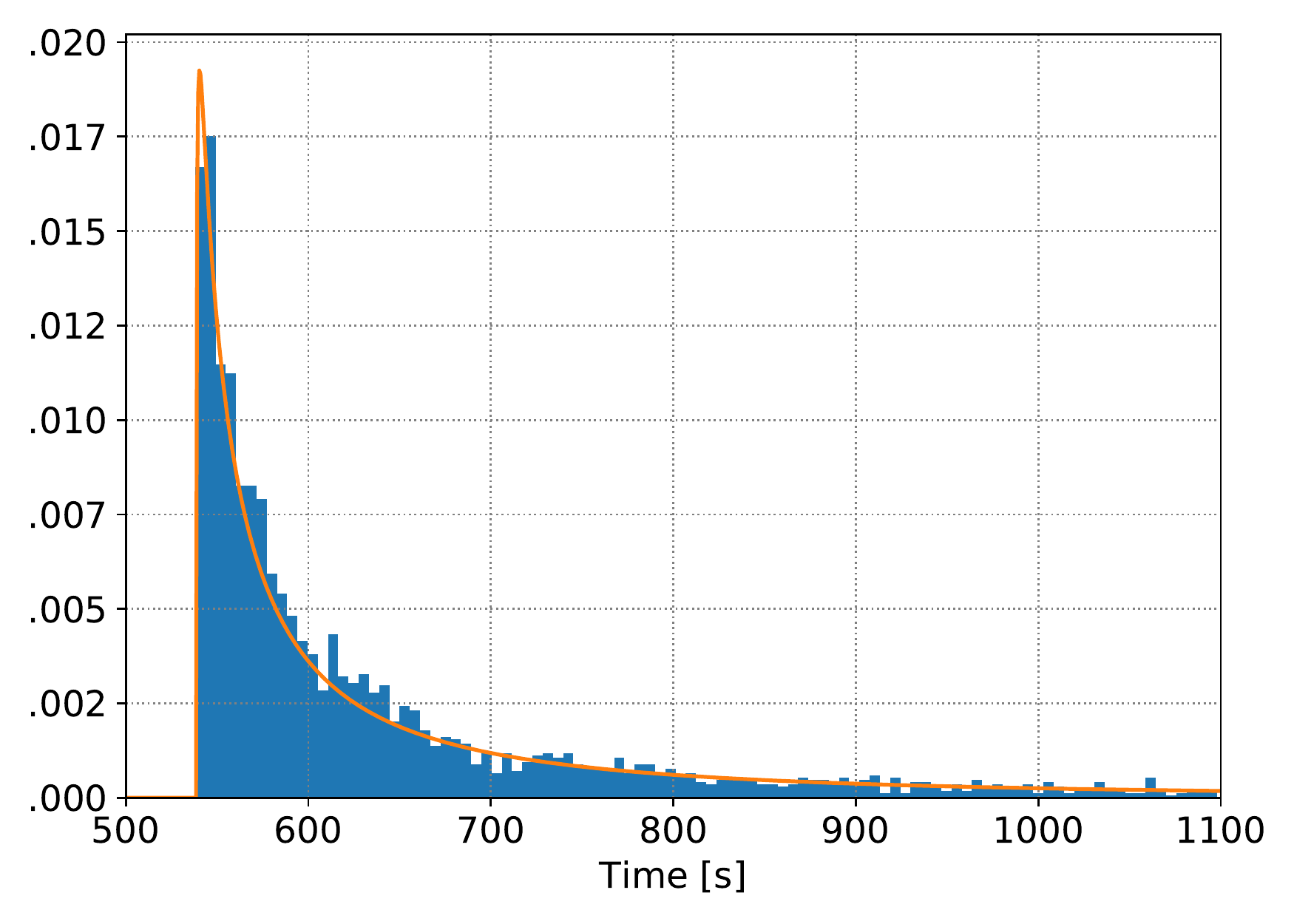}
  }
  \caption{Histograms for fault durations (a) and fault inter-arrival times (b) in the Antarex dataset compared to the PDFs of the Grid5000 data, as fitted on a Johnson SU and Exponentiated Weibull distribution respectively.}
  \label{fig:pdf}
 \end{figure}

FINJ orchestrates the execution of applications and the injection of faults by means of a workload file, as explained in Section~\ref{section:working}. For this purpose, we used several FINJ-generated workload files, one for each block of the dataset.

\paragraph{Workload Files} 
We used two different statistical distributions in the FINJ workload generator to create the durations and inter-arrival times of the tasks corresponding to the applications and fault-triggering programs. The application tasks are characterized by duration and inter-arrival times following normal distributions, and they occupy the 75\% of the dataset's entire duration. This resulted in tasks having an average duration of 30 minutes, and average inter-arrival times of nearly 40 minutes, both exhibiting relatively low variance and spread.

Fault-triggering tasks, on the other hand, are modeled using distributions fitted on the data from the failure trace associated to the Grid5000 cluster~\cite{bolze2006grid}, available on the Failure Trace Archive\footnote{\url{http://fta.scem.uws.edu.au/}}. We extracted from this trace the inter-arrival times of the host failures. Such data was then scaled and shifted to obtain an average of 10 minutes, while retaining the shape of the distribution. We then fitted the data using an exponentiated Weibull distribution, which is commonly used to characterize failure inter-arrival times~\cite{gainaru2015errors}. To model durations, we extracted for all hosts the time intervals between successive \emph{absent} and \emph{alive} records, which respectively describe connectivity loss and recovery events from the HPC system's resource manager to the host. We then fitted a Johnson SU distribution over a cluster of the data present at the 5 minutes point, which required no alteration in the original data. This particular distribution was chosen because of the quality of the fitting. In Figure~\ref{fig:pdf}, we show the histograms for the durations (a) and inter-arrival times (b) of the fault tasks in the workload files, together with the original distributions fitted from the Grid5000 data.

Unless configured to use specific probabilities, FINJ generates each task in the workload by randomly picking the respective application or fault program to be executed, from those that are available, with uniform probability. This implies that, statistically, all of the applications we selected will be subject to all of the available fault-triggering programs, given a sufficiently-long workload, with different execution overlaps depending on the starting times and durations of the specific tasks. Such a task distribution greatly mitigates overfitting issues. Finally, we do not allow fault-triggering program executions to overlap.

\paragraph{HPC Applications} 
We used a series of well-known benchmark applications, each of which stresses a different part of the node and measures the corresponding performance. 

\begin{enumerate}
\item DGEMM~\cite{dongarra1990algorithm}: performs CPU-heavy matrix-to-matrix multiplication;
\item HPC Challenge (HPCC)~\cite{luszczek2006hpc}: a collection of applications that stress both the CPU and memory bandwidth of an HPC system;
\item Intel distribution for High-Performance Linpack (HPL)~\cite{dongarra2003linpack}: solves a system of linear equations;
\item STREAM~\cite{stream}: stresses the system's memory and measures its bandwidth;
\item Bonnie++~\cite{bonnie}: performs HDD read-write operations;
\item IOZone~\cite{iozone}: performs HDD read-write operations. 
\end{enumerate}

The different applications provide a diverse environment for fault injection. Since we limit our analysis to a single machine, we use versions of the applications that rely on shared-memory parallelism, for example through the OpenMP library. 

\paragraph{Fault Programs}
All the fault programs used to reproduce anomalous conditions on Antarex are available at the FINJ Github repository. As described by Tuncer et al.~\cite{tuncer2018online}, each program can also operate in a low-intensity mode, thus doubling the number of possible faults. While we do not physically damage hardware, we closely reproduce several reversible hardware issues, such as I/O and memory allocation errors. Some of the fault programs (\emph{ddot} and \emph{dial}) only affect the performance of the CPU core they run on, whereas the others affect the entire compute node. The programs and the generated faults are as follows.

\begin{enumerate}
\item \emph{leak} periodically allocates 16MB arrays that are never released~\cite{tuncer2018online} creating a \emph{memory leak}, causing memory fragmentation and severe system slowdown;
\item \emph{memeater} allocates, writes into and expands a 36MB array~\cite{tuncer2018online}, decreasing performance through a \emph{memory interference} fault and saturating bandwidth;
\item \emph{ddot} repeatedly calculates the dot product between two equal-size matrices. The sizes of the matrices change periodically between 0.9, 5 and 10 times the CPU cache's size~\cite{tuncer2018online}. This produces a \emph{CPU and cache interference} fault, resulting in degraded performance of the affected CPU;
\item \emph{dial} repeatedly performs floating-point operations over random numbers~\cite{tuncer2018online}, producing an \emph{ALU interference} fault, resulting in degraded performance for applications running on the same core as the program;
\item \emph{cpufreq} decreases the maximum allowed CPU frequency by 50\% through the Linux Intel P-State driver.\footnote{\url{https://kernel.org/doc/Documentation/cpu-freq}} This simulates a \emph{system misconfiguration} or \emph{failing CPU} fault, resulting in degraded performance;
\item \emph{pagefail} makes any page allocation request fail with 50\% probability.\footnote{\url{https://kernel.org/doc/Documentation/fault-injection}} This simulates a \emph{system misconfiguration} or \emph{failing memory} fault, causing performance degradation and stalling of running applications;
\item \emph{ioerr} fails one out of 500 hard-drive I/O operations with 20\% probability, simulating a \emph{failing hard drive} fault, and causing degraded performance for I/O-bound applications, as well as potential errors;
\item \emph{copy} repeatedly writes and then reads back a 400MB file from a hard drive. After such a cycle, the program sleeps for 2 seconds~\cite{guan2013adaptive}. This simulates an \emph{I/O interference} or \emph{failing hard drive} fault by saturating I/O bandwidth, and results in degraded performance for I/O-bound applications.
\end{enumerate}

The faults triggered by these programs can be grouped in three categories according to their nature. The \emph{interference} faults (i.e., leak, memeater, ddot, dial and copy) usually occur when orphan processes are left running in the system, saturating resources and slowing down the other processes. \emph{Misconfiguration} faults occur when a component's behavior is outside of its specification, due to a misconfiguration by the users or administrators (i.e., cpufreq). Finally, the \emph{hardware} faults are related to physical components in the system that are about to fail, and trigger various kinds of errors (i.e., pagefail or ioerr). We note that some faults may belong to multiple categories, as they can be triggered by different factors in the system.

\section{Creation of Features}
\label{section:features}
In this section, we describe the process of parsing the metrics collected by LDMS to obtain a set of features capable of describing the state of the system, in particular for classification tasks.

\paragraph{Post-Processing of Data} 
Firstly, we removed all constant metrics (e.g., the amount of total memory in the node), which were redundant, and replaced the raw monotonic counters captured by the \emph{perfevent} and \emph{procinterrupts} plug-ins with their first-order derivatives. Moreover, we created an \emph{allocated} metric, both at the CPU core and node level, and integrated it in the original set. This metric can assume only binary values, and it encapsulates the information about the hardware component occupation, namely defining whether there is an application allocated on the node or not. This information would be available also in a real setting, since resource managers in HPC systems always keep track of the running jobs and their allocated computational resources. Lastly, for each above-mentioned metric and for each time point, we computed its first-order derivative and added it to the dataset, as proposed by Guan et al.~\cite{guan2012cda}.

After having post-processed the LDMS metrics, we created the feature sets via aggregation. Each feature set corresponds to a 60-second aggregation window and is related to a specific CPU core. The time step between consecutive feature sets is 10 seconds; this fine granularity allows for quick response times in fault detection. For each metric, we computed several statistical measures over the distribution of the values measured within the aggregation window~\cite{tuncer2018online}. These measures are the \emph{average}, \emph{standard deviation}, \emph{median}, \emph{minimum}, \emph{maximum}, \emph{skewness}, \emph{kurtosis}, and the \emph{5th, 25th, 75th} and \emph{95th percentiles}. Overall, there are 22 statistical features for each metric in the dataset (including also the first-order derivatives). Hence, starting from the initial set of 2094 LDMS metrics including node-level data as well as from all CPU cores, the final feature sets contain 3168 elements for each separate CPU core. We note that we did not include the metrics collected by the \emph{procinterrupts} plugin, as a preliminary analysis revealed them to be irrelevant for fault classification. All the scripts used to process the data are available on the FINJ Github repository.

\paragraph{Labeling} 
To train classifiers to distinguish between faulty and normal states, we labeled the feature sets either according to the fault program (i.e., one of the 8 programs presented in Section~\ref{subsection:features}) running within the corresponding aggregation window, or ``healthy'' if no fault was running. The logs produced by FINJ (included in the Antarex dataset) provide the details about the fault programs running at each time-stamp. In a realistic deployment scenario, the fault detection model can also be trained using data from spontaneous, real faults. In that case, administrators should provide explicit labels instead of relying on fault injection experiments.

A single aggregation window may capture multiple system states, making labeling not trivial. For example, a feature set may contain ``healthy'' time points before and after the start of a fault, or even include two different fault types. We define these feature sets as \emph{ambiguous}. One of the reasons behind the use of a short aggregation window (60 seconds) is to minimize the impact of such ambiguous system states on fault detection. However, since these situations cannot be avoided, we propose two labelling methods. The first method is \emph{mode}, where all the labels appearing in the time window are considered. Their distribution is examined and the label with the majority of occurrences is assigned to the feature set. This leads to robust feature sets, whose labels are always representative of the aggregated data. The second method is \emph{recent}, where the label is obtained by observing the state of the system at the most recent time point in the time window (the last time point). This could correspond to a certain fault type or could be ``healthy''. This approach could be less robust, for instance when a feature set that is mostly ``healthy'' is labelled as faulty, but has the advantage of leading to a more responsive fault detection, as faults are detected as soon as they are encountered, without having to look at the state over the last 60 seconds.

\paragraph{Detection System Architecture}

\begin{figure}[t]
 \centering
  \includegraphics[width=0.45\textwidth,trim={0 0 0 0}, clip=true]{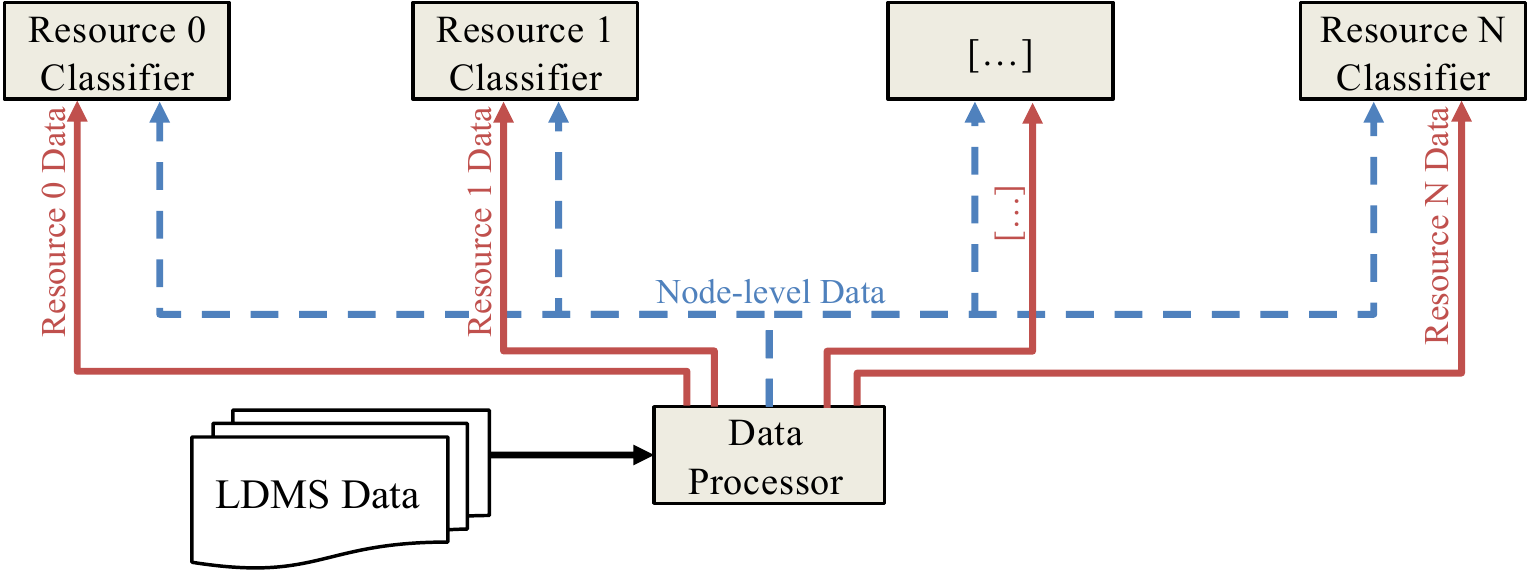}
  \caption{Architecture of the proposed machine learning-based fault detection system.}
  \label{fig:detarchitecture}
 \end{figure}

The fault detection system we propose in this paper is based on an architecture containing an array of classifiers, as shown in Figure~\ref{fig:detarchitecture}. A classifier handles a specific computing resource type in the node, such as a CPU core, GPU or MIC. Each classifier is trained with the feature sets of all the resource units of the corresponding type, and is able to perform fault diagnoses for all of them, thus detecting faults both at node level and resource level (e.g., dial and ddot). This is possible as the feature sets of each classifier contain all \emph{node-level} metrics for the system as well as the \emph{resource-specific} metrics for the resource unit. Since a feature set contains data from only one resource type, the proposed approach allows us to limit their size, which in turn improves performance and detection accuracy. The key assumption of this approach is that the resource units of the same type behave similarly and that the respective feature sets can be combined in a coherent set. It is anyway possible to use a separate classifier for each resource unit of the same type without altering the feature sets themselves, should this assumption prove to be too strong. In our case, the computing nodes have only CPU cores. Therefore, we train one classifier with feature sets that contain both node-level and core-level data, for one core at a time.

The classifiers' training can be performed offline, using labeled data resulting from normal system operation or from fault injection, as we do in this paper. The trained classifiers can then be deployed to detect faults on new, streaming data. By design, our architecture can detect only one fault at any time. If two faults happen simultaneously, the classifier would detect the fault whose effects on the system are deemed more disruptive for the normal, ``healthy'' state. In this preliminary study, this design choice is reasonable, as our purpose is to automatically distinguish between different fault conditions. Although unlikely, multiple faults could affect the same compute node of an HPC system at the same time. Our approach could be extended to deal with this situation by devising a classifier that does not produce a single output but rather a composite output, for instance a distribution or a set of probabilities, one for each possible fault type.

\section{Experimental Results}
\label{section:experimentalresults}

In this section we present the results of our experiments, whose purpose was to correctly detect which of the 8 faults (as described in Section~\ref{subsection:features}) were injected in the Antarex HPC node at any point in time in the Antarex dataset. Along the way, we compared a variety of classification models and the two labeling methods discussed in Section~\ref{section:features}, assessed the impact of ambiguous feature sets, estimated the most important metrics for fault detection, and finally evaluated the overhead of our detection method. For each classifier, both the training and test sets are built from the feature set of one randomly-selected core for each time point. We evaluated the classifiers using 5-fold cross-validation, with the average F-score as the performance metric.\footnote{The F-score is the harmonic mean between precision and recall.} The software environment we used is Python 3.4 with the Scikit-learn package~\cite{pedregosa2011scikit}.

\begin{figure}[t!]
 \centering
 \captionsetup[subfigure]{}
  \subfloat[Time-stamp order.]{
    \includegraphics[width=0.45\textwidth,trim={0 0 0 0}, clip=true]{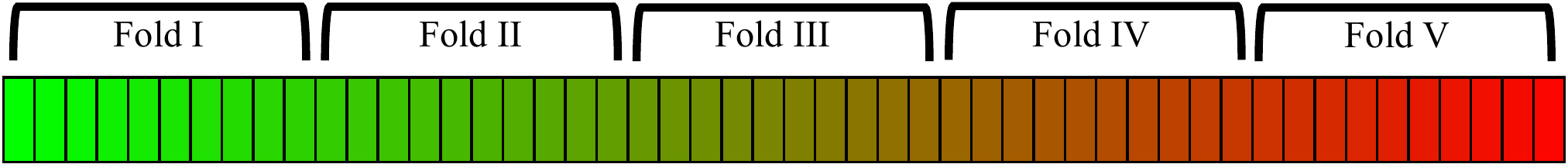}
  	}
  \\
    \subfloat[Shuffled order.]{
    \includegraphics[width=0.45\textwidth,trim={0 0 0 0}, clip=true]{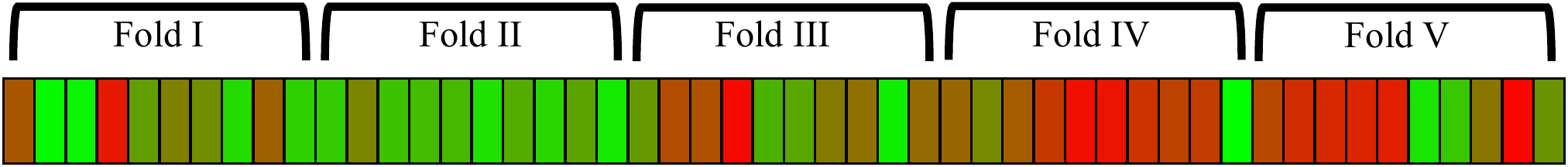}
  	}
  \caption{The effect of using time-stamp (a) or shuffled (b) ordering to create the data folds for cross-validation. Blocks with similar color represent feature sets from the same time frame.}
  \label{fig:shuffling_folds}
 \end{figure}

Although data shuffling is a standard technique with proven advantages in machine learning, it is not well suited to the fault detection method we propose in this paper. This is because our design is tailored for online systems, where classifiers are trained using only continuous, streamed, and potentially unbalanced data as it is acquired, while ensuring robustness in training so as to detect faults in the near future. Hence, it is very important to assess the detection accuracy without data shuffling. We reproduced this realistic, online scenario by performing cross-validation on the Antarex dataset using feature sets in time-stamp order. Time-stamp ordering produces cross-validation folds that contain data from specific time intervals. We depict this effect in Figure~\ref{fig:shuffling_folds}. In any case, we used shuffling in a small subset of the experiments for comparative purposes.

\subsection{Comparison of Classification Models}
\label{section:classifiers}

We first present results associated with different classifiers using feature sets in time-stamp order and the \emph{mode} labeling method. As classification models, we opted for a Random Forest (RF), Decision Tree (DT), Linear Support Vector Classifier (SVC) and Neural Network (NN) with two hidden layers, each having 1000 neurons. A preliminary empirical evaluation revealed that this neural network architecture, as well as the other models, are well-suited for our purposes and thus provide good results. Even though we also considered using more complex models, such as \emph{Long Short Term Memory} (LSTM) neural networks, we ultimately decided to employ comparable, general-purpose models. This allows us to broaden the scope of our study, evaluating the impact of factors such as data normalization, shuffling and ambiguous system states on fault classification. On the other hand, LSTM-like models have more restrictive training requirements and are more tailored for regression tasks. Finally, since the Antarex dataset was acquired by injecting faults lasting a few minutes each, such a model would not benefit from the long-term correlations in system states, where models like RF are capable of near-optimal performance.

\begin{figure}[t!]
 \centering
 \captionsetup[subfigure]{}
  \subfloat[Random Forest.]{
    \includegraphics[width=0.45\textwidth,trim={0 60 0 0}, clip=true]{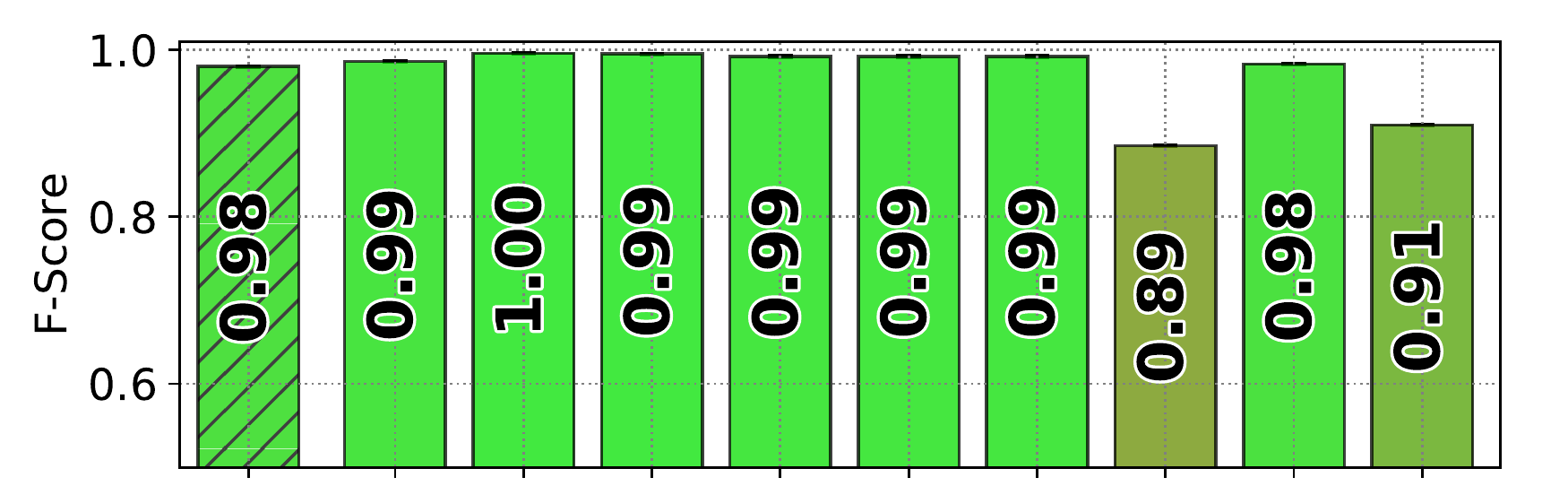}
  	} \\
  \subfloat[Decision Tree.]{
    \includegraphics[width=0.45\textwidth,trim={0 60 0 0}, clip=true]{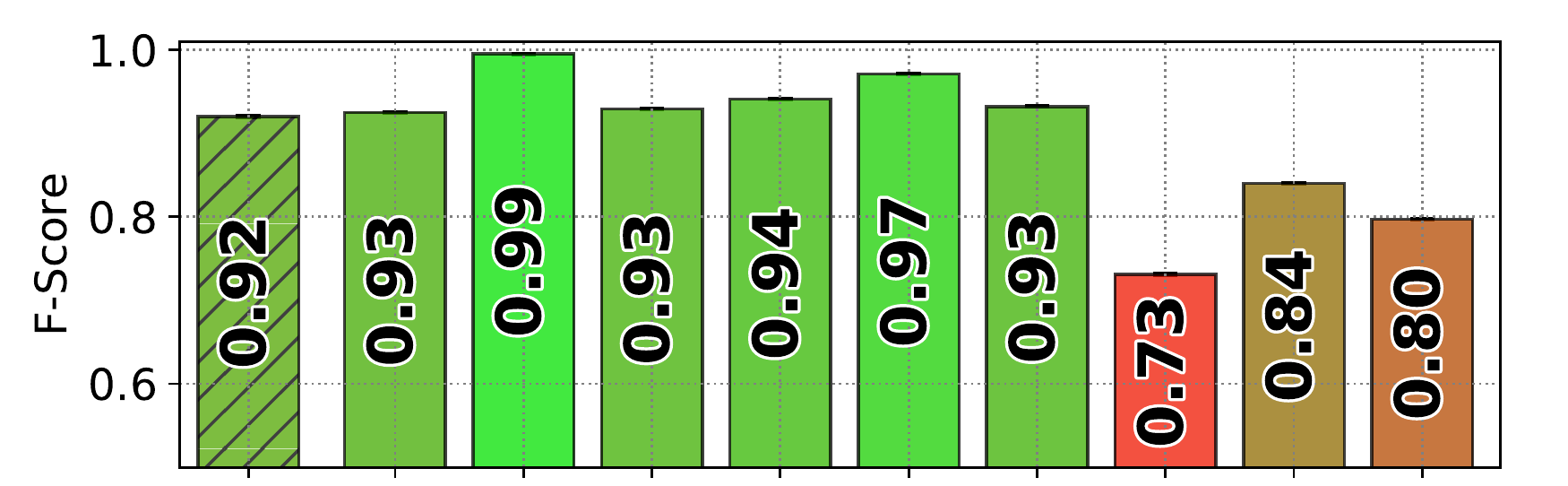}
  }
  \\
    \subfloat[Neural Network.]{
    \includegraphics[width=0.45\textwidth,trim={0 60 0 0}, clip=true]{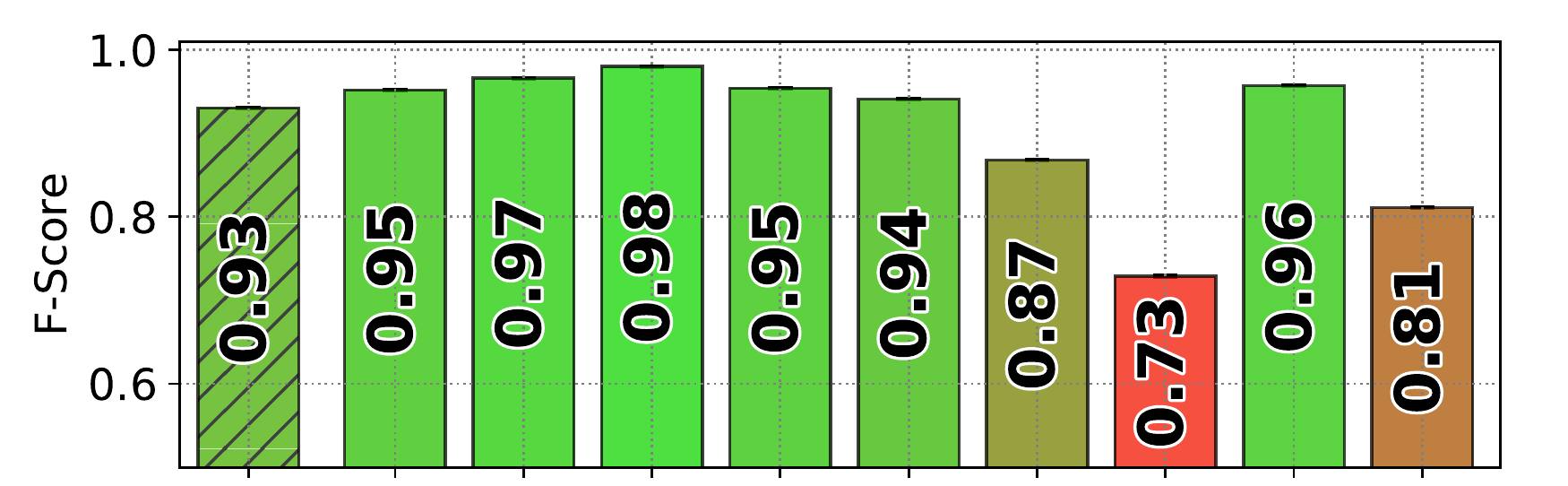}
  	} \\
  \subfloat[Support Vector Classifier.]{
    \includegraphics[width=0.45\textwidth,trim={0 0 0 0}, clip=true]{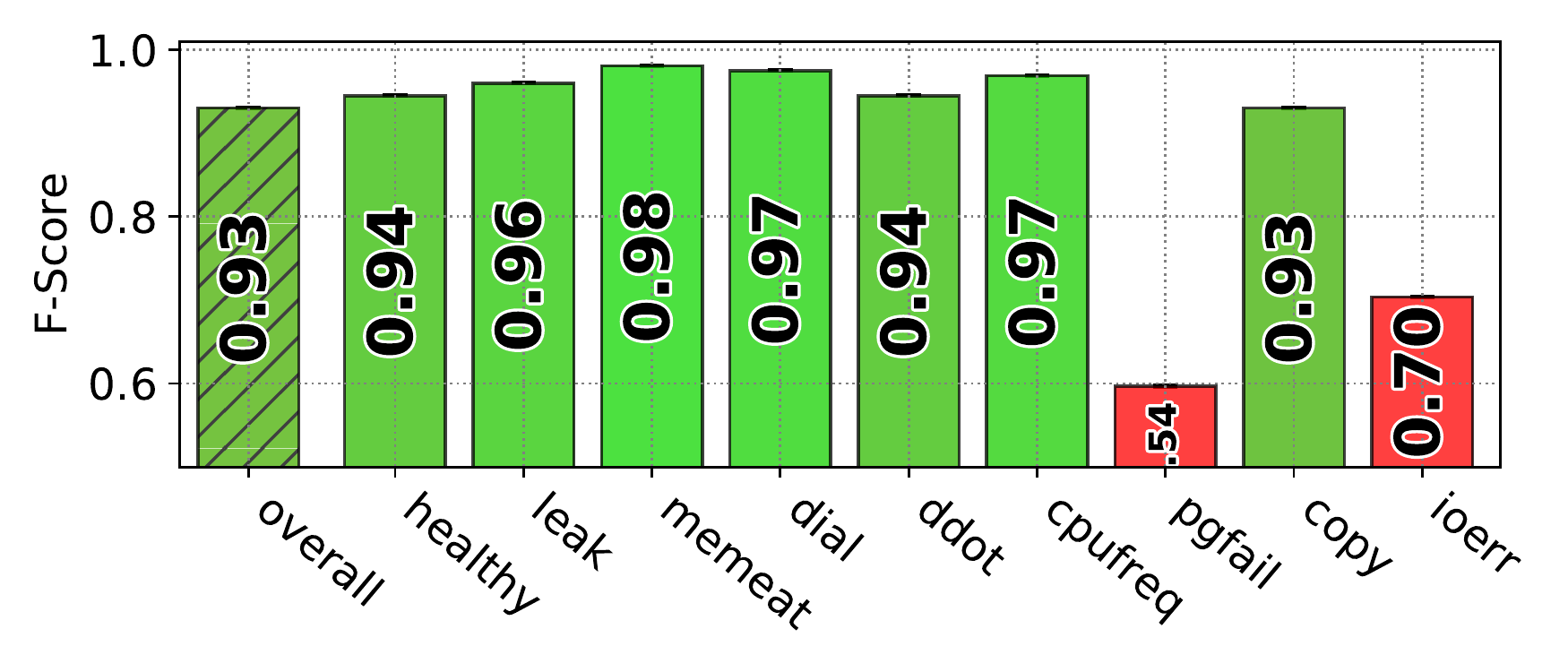}
  }
  \caption{Classification results on the Antarex dataset, using all feature sets in time-stamp order, the \emph{mode} labeling method, and different classification models.}
  \label{fig:classifiers}
 \end{figure}

The results of each classifier and each fault type are shown in Figure~\ref{fig:classifiers}, with the overall F-score highlighted. We notice that all classifiers have very good performance, with the overall F-scores well above 0.9. RF is the best classifier, with an overall F-score of 0.98, followed by NN and SVC scoring 0.93. The most difficult fault types to detect for all classifiers are \emph{pagefail} and \emph{ioerr} faults, which have substantially worse scores than the others. 

We infer from the results that an RF would be the ideal classifier for an online fault detection system, due to its detection accuracy which is at least 5\% higher than the other classifiers, in terms of the overall F-score. Additionally, random forests are computationally efficient~\cite{lakshminarayanan2014mondrian}, and therefore would be suitable for use in online environments with strict overhead requirements. As an additional advantage, unlike the NN and SVC classifiers, RF and DT do not require data normalization. Normalization in an online environment is hard to achieve, as many metrics do not have well-defined upper bounds. Although a rolling window-based dynamic normalization approach can be used~\cite{guan2013adaptive} to solve the problem, this method is unfeasible for ML-based classification, as it can lead to quickly-degrading detection accuracy and to the necessity of frequent training. For all these reasons, we will show only the results of an RF classifier in the following. 

\subsection{Comparison of Labeling Methods and Shuffling}
 
\begin{figure}[t!]
 \centering
 \captionsetup[subfigure]{}
    \subfloat[Mode labeling.]{
    \includegraphics[width=0.45\textwidth,trim={0 60 0 0}, clip=true]{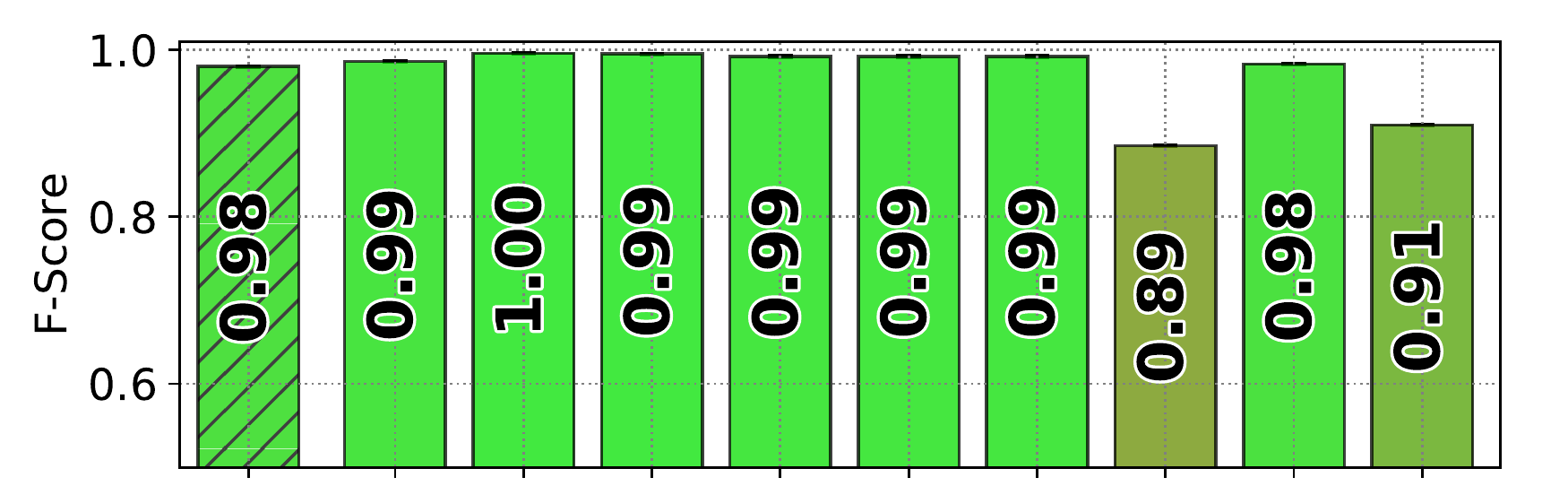}
  	} \\
  \subfloat[Recent labeling.]{
    \includegraphics[width=0.45\textwidth,trim={0 60 0 0}, clip=true]{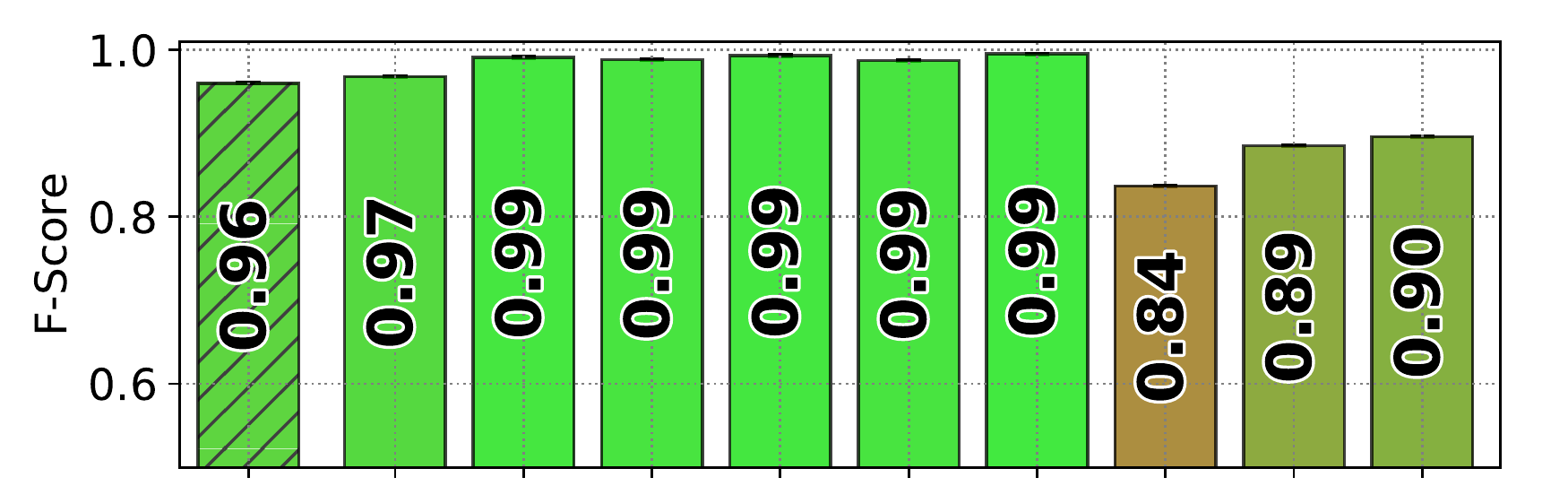}
  }
  \\
    \subfloat[Mode labeling with shuffling.]{
    \includegraphics[width=0.45\textwidth,trim={0 60 0 0}, clip=true]{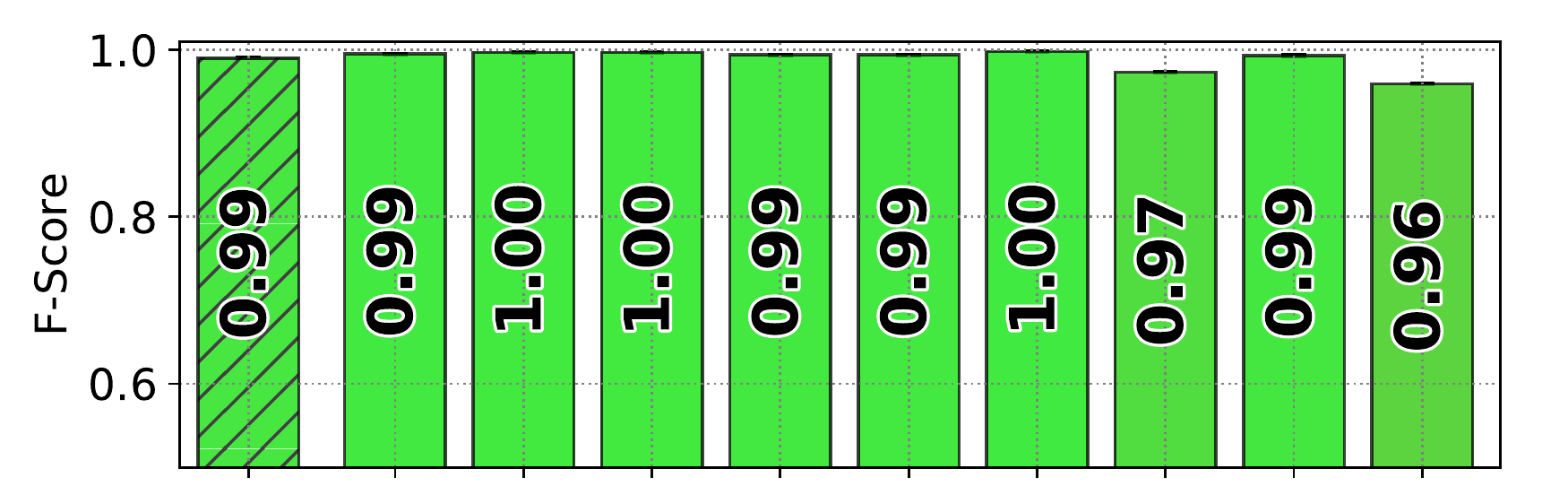}
  	} \\
  \subfloat[Recent labeling with shuffling.]{
    \includegraphics[width=0.45\textwidth,trim={0 0 0 0}, clip=true]{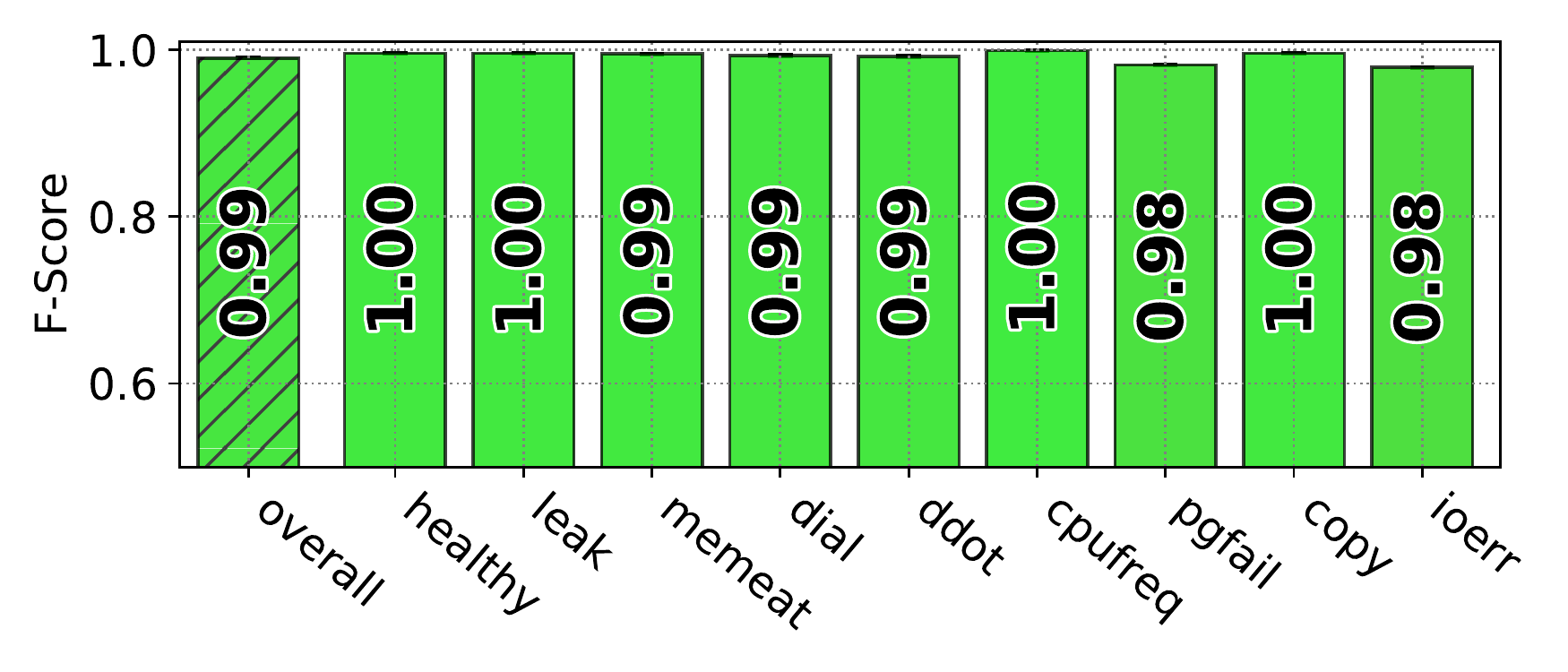}
  }
  \caption{RF classification results, using all feature sets in time-stamp or shuffled order, with the \emph{mode} and \emph{recent} labeling methods.}
  \label{fig:labeling}
 \end{figure}

Next we present the results of the two labeling methods described in Section~\ref{subsection:features}. Figures~\ref{fig:labeling}a and~\ref{fig:labeling}b report the classification results without data shuffling for, respectively, the \emph{mode} and the \emph{recent} labeling. The overall F-scores are 0.98 and 0.96, close to the optimal values. Once again, in both cases the \emph{ioerr} and \emph{pagefail} faults are substantially more difficult to detect than the others. This is probably due to the intermittent nature of both of these faults, whose effects depend on the hard drive I/O (ioerr) and memory allocation (pagefail) patterns of the underlying applications.

We observe an interesting behavior with the \emph{copy} fault program, which gives a worse F-score when using the \emph{recent} method in~\ref{fig:labeling}b. As shown in Section~\ref{section:metrics}, a metric related to the read rate of the hard drive used in our experiments (\emph{time\_read\_rate\_der\_perc95}) is defined as important by the DT model for distinguishing faults, and we assume it is useful for detecting hard drive faults in particular, since it has no use for CPU-related faults. However, this is a comparatively slowly-changing metric. For this reason, a feature set may be labeled as \emph{copy} as soon as the program is started, before the metric has been updated to reflect the new system status. This in turn makes classification more difficult and leads to degraded accuracy. This leads us to conclude that \emph{recent} may not be well suited for the faults whose effects cannot be detected immediately.

Figures~\ref{fig:labeling}c and~\ref{fig:labeling}d report the detection accuracy for the \emph{mode} and \emph{recent} methods, this time obtained after having shuffled the data for the classifier training phase. As expected, data shuffling markedly increases the detection accuracy for both labeling methods, reaching an almost optimal F-score with all fault types -- and overall F-score of 0.99. A similar performance boost with data shuffling was obtained also with the remaining classification models introduced in Section~\ref{section:classifiers} (the results are not reported here since they would not add any insight to the experimental analysis). We notice that the \emph{recent} labeling has a slightly higher detection rate, especially for some fault types. The reason for this improvement most likely lies in the highly reactive nature of this labeling method, as it can capture status changes faster than \emph{mode}. Another interesting observation is that adding data shuffling grants a larger performance boost to the \emph{recent} labeling compared to the \emph{mode} labeling. This happens because the \emph{recent} method is more sensible to temporal correlations in the data, which in turn may lead to classification errors. Data shuffling destroys the temporal correlations in the training set and thus improves detection accuracy. 

\subsection{Impact of Ambiguous Feature Sets}

\begin{figure}[t!]
 \centering
 \captionsetup[subfigure]{}
    \subfloat[Non-ambiguous dataset.]{
    \includegraphics[width=0.45\textwidth,trim={0 60 0 0}, clip=true]{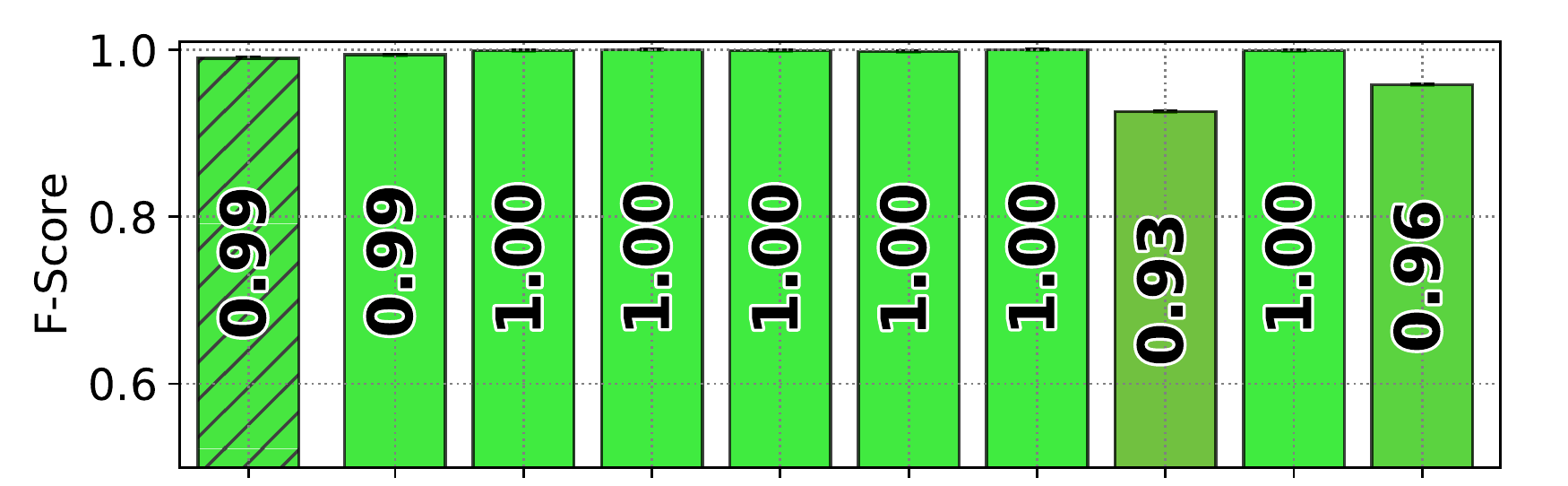}
  	} \\
  \subfloat[Non-ambiguous dataset with shuffling.]{
    \includegraphics[width=0.45\textwidth,trim={0 0 0 0}, clip=true]{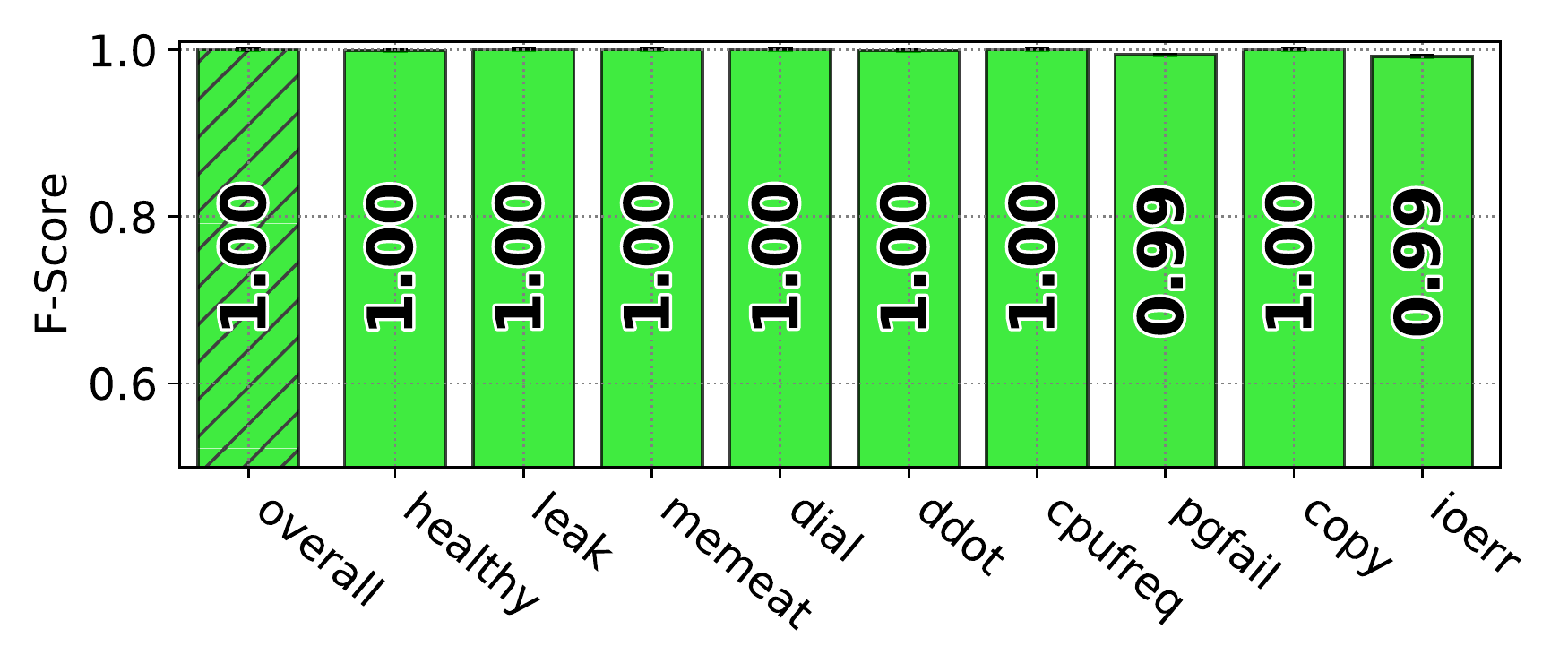}
  }
  \caption{RF classification results on the Antarex dataset, using only non-ambiguous feature sets in time-stamp (a) and shuffled (b) order.}
  \label{fig:nonambiguous}
 \end{figure}

Here we give insights on the impact of ambiguous feature sets in the dataset on the classification process by excluding them from the training and test sets. As shown in Figure~\ref{fig:nonambiguous}, the overall F-scores are above 0.99 both without (Figure~\ref{fig:nonambiguous}a) and with shuffling (Figure~\ref{fig:nonambiguous}b), leading to a slightly better classification performance compared to having the ambiguous feature sets in the dataset. Around 20\% of the feature sets of the Antarex dataset is ambiguous. With respect to this relatively large proportion, the performance gap described above is small, which proves the robustness of our detection method. In general, the proportion of ambiguous feature sets in a dataset depends primarily on the length of the aggregation window, and on the frequency of state changes in the HPC system. More feature sets will be ambiguous as the length of the aggregation window increases, leading to more pronounced adverse effects on the classification accuracy. Thus, as a practical guideline, we advise to use a short aggregation window, such as the 60-second window we employed here, to perform online fault detection. 

\begin{figure*}[t!]
 \centering
 \captionsetup[subfigure]{}
    \subfloat[ddot.]{
    \includegraphics[width=0.32\textwidth,trim={0 0 0 0}, clip=true]{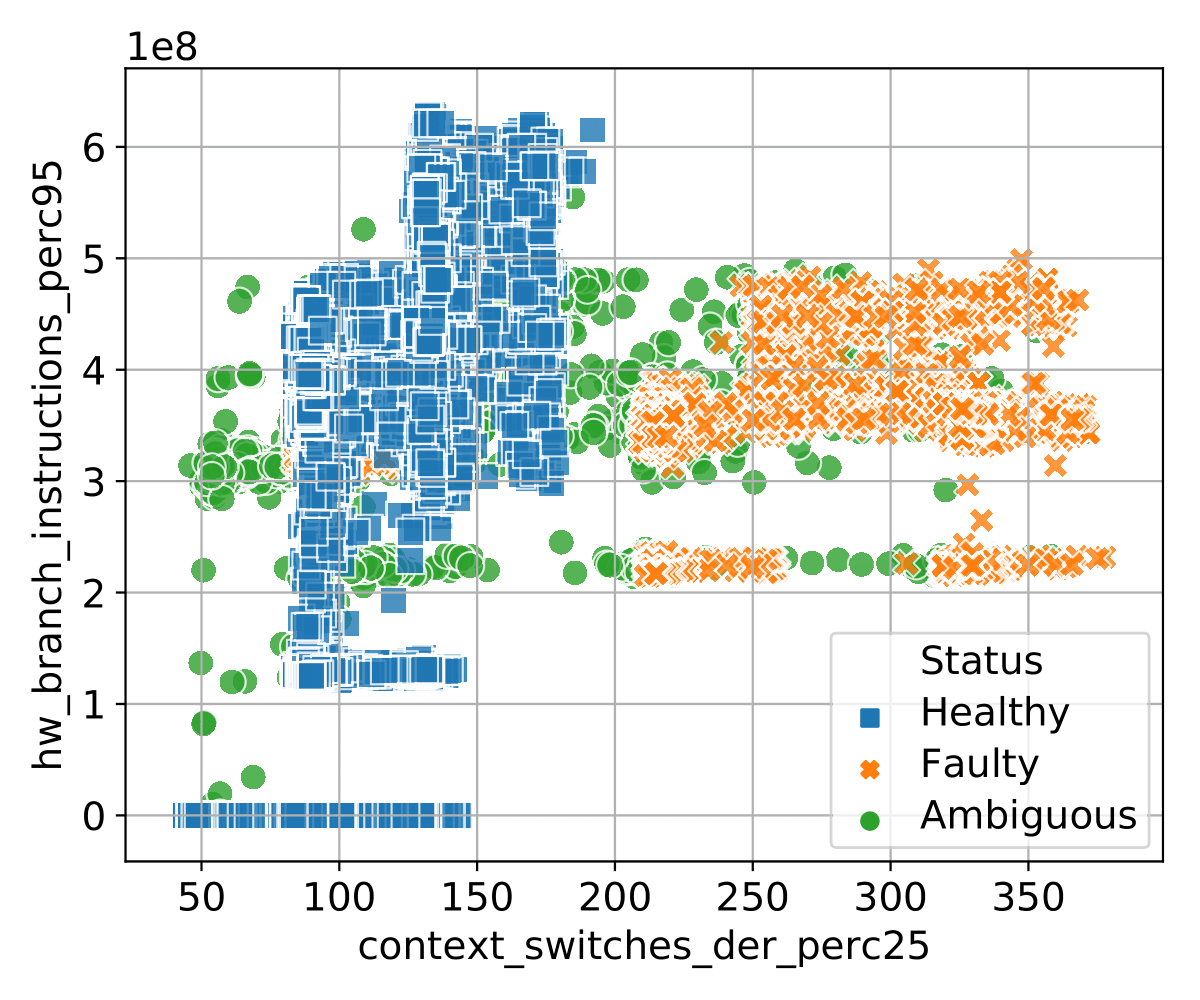}
  	}
  \subfloat[cpufreq.]{
    \includegraphics[width=0.32\textwidth,trim={0 0 0 0}, clip=true]{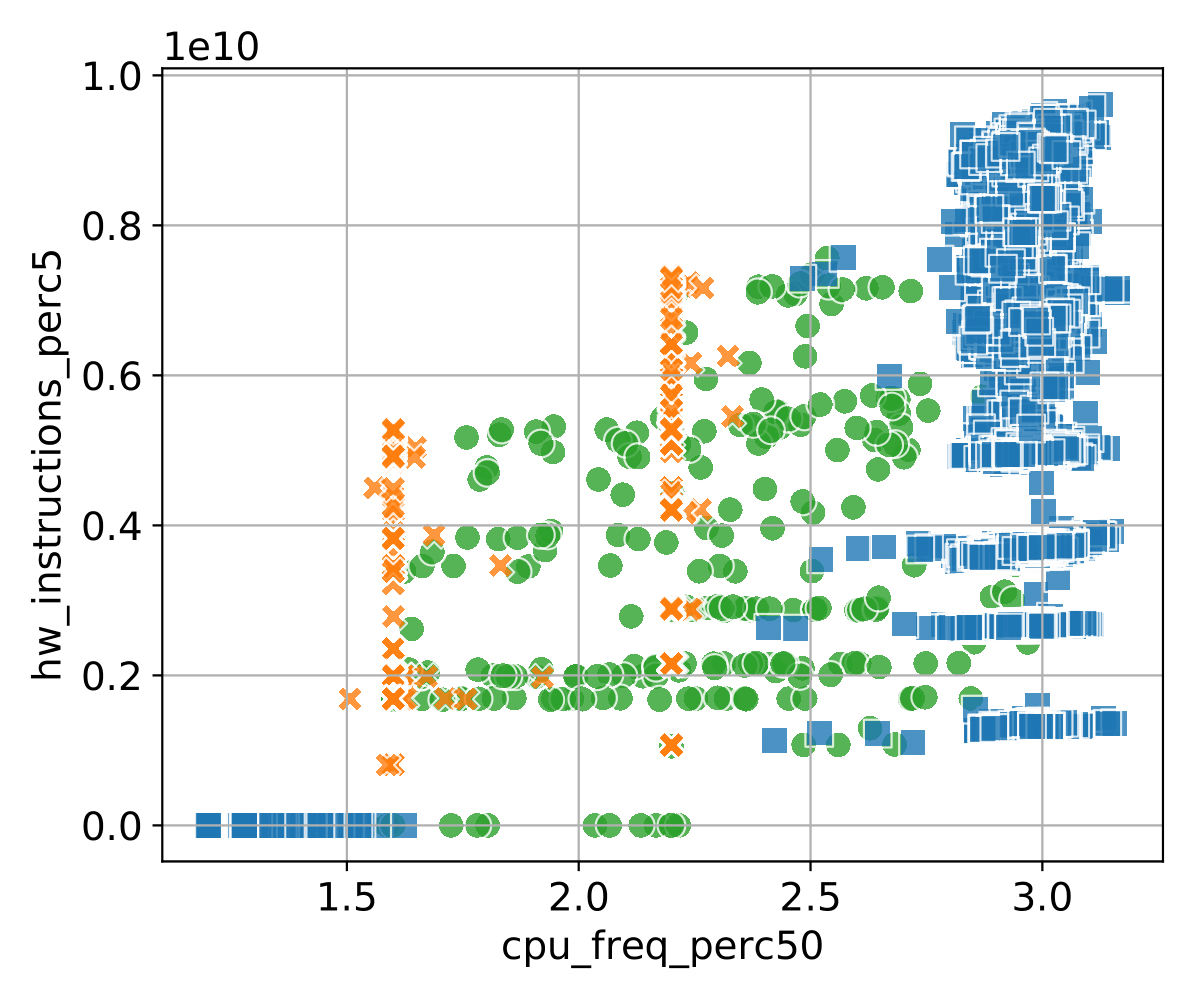}
  }
  \subfloat[memeater.]{
    \includegraphics[width=0.31\textwidth,trim={0 0 0 0}, clip=true]{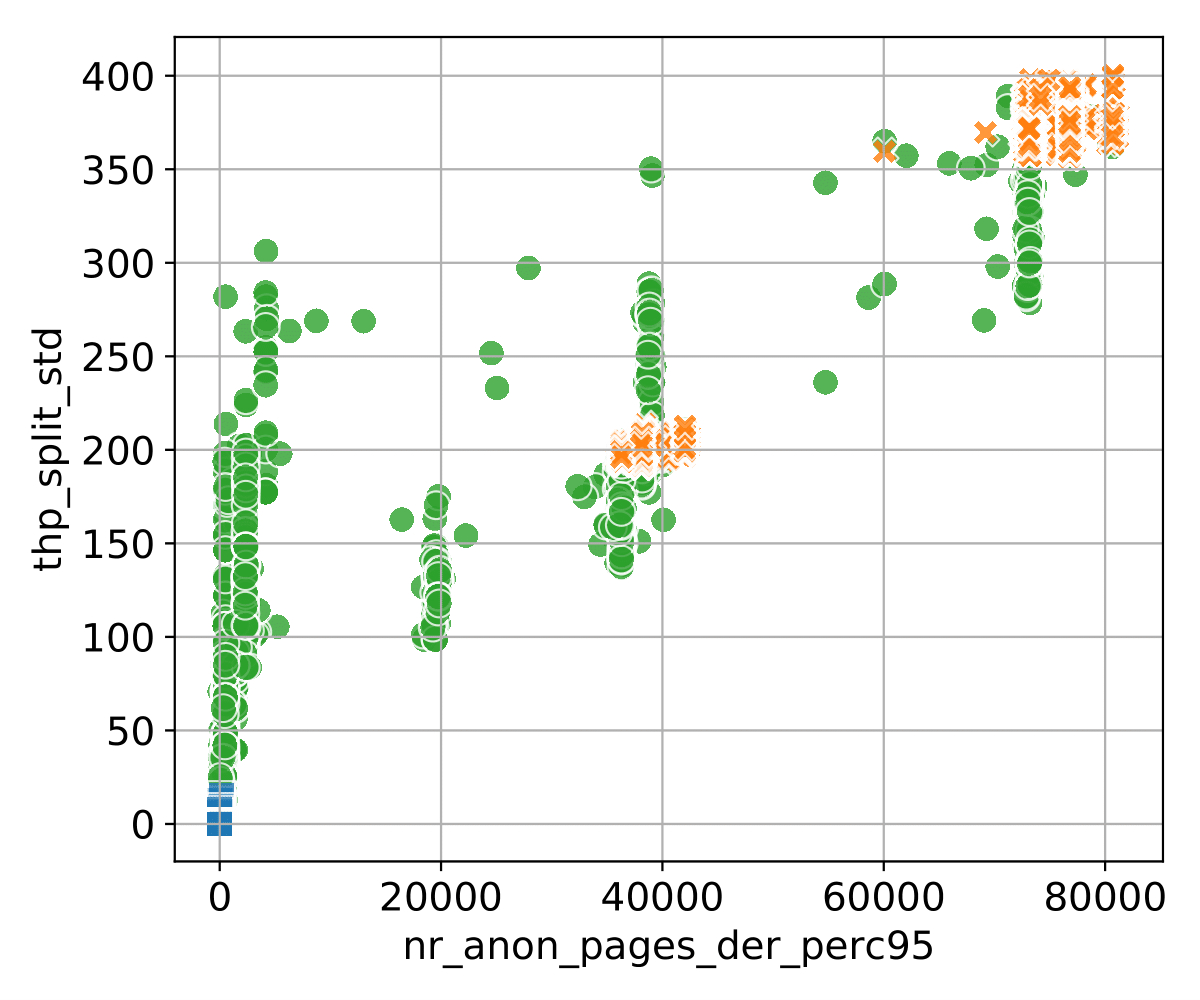}
  }
  \caption{The scatter plots of two important metrics, as quantified by a DT classifier, for three fault types. The ``healthy'' points are marked in blue, while fault-affected points in orange, and the points related to ambiguous feature sets in green.}
  \label{fig:scatter_plots}
 \end{figure*}

A more concrete example of the behavior of ambiguous feature sets can be seen in Figure~\ref{fig:scatter_plots}, where we show the scatter plots of two important metrics (as we will discuss in Section~\ref{section:metrics}) for the feature sets related to the \emph{ddot}, \emph{cpufreq} and \emph{memeater} fault programs, respectively. The ``healthy'' points, marked in blue, and the fault-affected points, marked in orange, are distinctly clustered in all cases. On the other hand, the points representing the ambiguous feature sets, marked in green, are sparse, and often fall right between the ``healthy'' and faulty clusters. This is particularly evident with the cpufreq fault program in Figure~\ref{fig:scatter_plots}b.

\subsection{Estimation of the Most Important Metrics}
 \label{section:metrics}

A crucial aspect in fault detection is understanding the most important metrics for the detection accuracy. Identifying them can help reducing the amount of collected data, thus reducing the number of hardware measuring components or software tools which could create additional overhead. Tuncer et al.~\cite{tuncer2018online} showed that using methods such as principal component analysis (and other methods that rely exclusively on the variance in the data) may discard certain important metrics. On the other hand, a RF classifier tends to report as relevant the same metric many times, with different statistical indicators. This is caused by its ensemble nature (a random forest is comprised of a collection of decision trees) and the subtle differences in the estimators that compose it. Instead, a DT classifier naturally provides this information, as the most relevant metrics will be those in the top layer of the decision tree.\footnote{Decision trees are built by splitting the data in subsets. The splitting choice is based on the value of the metrics, or attribute in the DT terminology. The attributes providing the highest information gain (i.e., the most relevant one) are selected first by the standard DT training algorithms.} Thus, we trained a DT classifier on the Antarex dataset.

The results are listed in Table~\ref{table:imp_metrics}, along with their source LDMS plug-ins. While the metrics marked in bold are per-core, the others are system-wide. We notice that metrics from most of the available plug-ins are used, and some of these metrics can be directly associated to the behaviour of the faults. For instance, the metric related to context switches (\emph{context\_switches\_der\_perc25}) is tied to the \emph{dial} and \emph{ddot} programs, as CPU interference generates an anomalous number of context switches. In general, first-order derivatives (marked with the ``der'' suffix) are widely used by the classifier, which suggests that computing them is actually useful for fault detection. On the contrary, more complex statistical indicators such as the skewness and kurtosis do not appear among the most relevant. This suggests that simple features are sufficient for machine learning-driven fault detection on HPC nodes.

\begin{table}[t]
\caption{The most important metrics for fault detection, obtained via a DT classifier.}
\label{table:imp_metrics}
\centering
\fontsize{9.5}{9.5}\selectfont
\begin{tabular}{l|l}
Source & Name   \\ \hline
\texttt{procsensors} & 1. \textbf{cpu\_freq\_perc50}  \\
\texttt{meminfo} & 2. active\_der\_perc5 \\
\texttt{perfevent} & 3. \textbf{hw\_cache\_misses\_perc50} \\
\texttt{vmstat} & 4. thp\_split\_std  \\
\texttt{vmstat} & 5. nr\_active\_file\_der\_perc25 \\
\texttt{perfevent} & 6. \textbf{hw\_branch\_instructions\_perc95} \\
\texttt{meminfo} & 7. mapped\_der\_avg  \\
\texttt{meminfo} & 8. nr\_anon\_pages\_der\_perc95  \\
\texttt{procstat} & 9. sys\_der\_min \\
\texttt{vmstat} & 10. nr\_dirtied\_der\_std \\ 
\texttt{meminfo} & 11. kernelstack\_perc50 \\
\texttt{vmstat} & 12. numa\_hit\_perc5  \\
\texttt{procstat} & 13. processes\_der\_std \\
\texttt{procstat} & 14. context\_switches\_der\_perc25 \\
\texttt{procstat} & 15. procs\_running\_perc5 \\
\texttt{finj} & 16. \textbf{allocated\_perc50} \\
\texttt{vmstat} & 17. nr\_free\_pages\_der\_min \\
\texttt{diskstats} & 18. time\_read\_rate\_der\_perc95 \\
\texttt{vmstat} & 19. nr\_kernel\_stack\_der\_max \\
\texttt{perfevent} & 20. \textbf{hw\_instructions\_perc5} 
\end{tabular}
\end{table}

\subsection{Remarks on Overhead}
Finally, a critical consideration for understanding the feasibility of a fault detection approach is its overhead, especially if the final target is its deployment in a real HPC system. LDMS is proven to have a low overhead at high sampling rates~\cite{agelastos2014lightweight}. We also assume that the generation of feature sets and the classification are performed locally in each node (\emph{on-edge} computation), and that only the resulting fault diagnoses are sent externally, which greatly decreases the network communication requirements and overhead. Following these assumptions, the hundreds of performance metrics used to train the classification models do not need to be sampled and streamed at a fine granularity. Generating a set of feature sets, one for each core in the test node, at a given time point for an aggregation window of 60 seconds requires, on average, 340 ms by employing a single thread. This time includes the I/O overhead due to reading and parsing the LDMS CSV files, and writing the output feature sets. RF classifiers are very efficient: classifying a single feature set as faulty or not requires 2ms, on average. This means that in total 342 ms are needed to generate and classify a feature set, using a single thread and a 60-seconds aggregation window. This is more than acceptable for online use and practically negligible. Furthermore, the overhead should be much lower in a real HPC system, with direct in-memory access to streamed data, as a significant fraction of the overhead in our simulation is due to file system I/O operations to access the CSV files with the data. Additionally, the single statistical features are independent from each other. This means that the data can be processed in parallel fashion, using multiple threads to further reduce latency and ensure load balancing across CPU cores, which is a critical aspect to prevent application slowdown induced by fault detection.
\section{Conclusions}
\label{section:conclusions}
We studied a ML approach to online fault classification in HPC systems. Our study provided three contributions to the state-of-the-art in resiliency research in the HPC systems field. The first is FINJ, a fault injection tool, which allows for the automation of complex experiments, and for reproducing anomalous behaviors in a deterministic, simple way. FINJ is implemented in Python and has no dependencies for its core operation. This, together with the simplicity of its command-line interface, makes the deployment of FINJ on large-scale systems trivial. Since FINJ is based on the use of tasks, which are external executable programs, users can integrate the tool with any existing lower-level fault injection framework that can be triggered in such way, and ranging from the application level to the kernel, or even hardware level. The use of workloads in FINJ also allows to reproduce complex, specific fault conditions, and to reliably perform experiments involving multiple nodes at the same time.

The second contribution is the Antarex fault dataset, which we generated using FINJ, to enable training and evaluation of our supervised ML classification models, and which we extensively described. Both FINJ and the Antarex dataset are publicly available to facilitate resiliency research in the HPC systems field. The third contribution is a classification method intended for streamed, online data obtained from a monitoring framework, which is then processed and fed to classifiers. The experimental results of our study show almost perfect classification accuracy for all fault types injected by FINJ, with negligible computational overhead for HPC nodes. Moreover, our study reproduces the operating conditions that could be found in a real online system, in particular those related to ambiguous system states and data imbalance in the training and test sets.

As future work, we plan to deploy our fault detection method in a large-scale real HPC system. This will involve facing a number of new challenges. We need to develop tools to aid online training of machine learning models, as well as integrate our method in a monitoring framework such as Examon~\cite{beneventi2017continuous}. We also expect to study our approach in online scenarios and adapt it where necessary. For instance, we need to characterize the scalability of FINJ, and extend it to include the ability to build workloads where the order of tasks is defined by \emph{causal} relationships rather than time-stamps, which might simplify the triggering of extremely specific anomalous states in a given system. Since training is performed before HPC nodes move into production (e.g., in a test environment), we also need to characterize how often re-training is needed, and devise a procedure to perform this. Finally, we plan to make our detection method robust against the occurrence of faults that were unknown during the training phase, preventing their misclassification, as well as expect to evaluate some specialized models such as LSTM neural networks, in the light of the general results obtained with this study.

\paragraph{Acknowledgements} A. Netti has been supported by the EU project \textit{Oprecomp-Open Transprecision Computing} (grant agreement 732631). A. S\^irbu has been partially funded by the EU project \textit{SoBigData Research Infrastructure --- Big Data and Social Mining Ecosystem} (grant agreement 654024). We thank the Integrated Systems Laboratory of ETH Zurich for granting us control of their Antarex HPC node during this study.

\bibliographystyle{abbrv}
\bibliography{bib}

\end{document}